\documentclass[journal,twoside,final]{IEEEtran}

\usepackage{comment}
\usepackage{url}

\usepackage{times}
\usepackage{epsfig}
\usepackage{graphicx}
\usepackage{amsmath}
\usepackage{amssymb}
\usepackage{amsthm}
\usepackage{enumerate}
\usepackage{indentfirst}
\usepackage{subfigure}
\usepackage{algorithmic}
\usepackage{algorithm}
\usepackage{url}
\usepackage{booktabs}
\usepackage{multirow}
\usepackage{longtable}
\usepackage{cite}

\theoremstyle{remark}
\newtheorem{theorem}{Theorem}

\newtheorem{lemma}{Lemma}

\long\def\symbolfootnote[#1]#2{\begingroup\def\thefootnote{\fnsymbol{footnote}} \footnote[#1]{#2}\endgroup}

\ifCLASSINFOpdf
\else
\fi
\usepackage{flushend}
\begin{document}

%
\title{Joint Beamforming and Power Control in Coordinated Multicell: Max-Min Duality, \\Effective Network and Large System Transition}
%
%
%

\author{Yichao Huang, \IEEEmembership{Member, IEEE}, Chee Wei Tan, \IEEEmembership{Senior Member, IEEE}, and Bhaskar D. Rao, \IEEEmembership{Fellow, IEEE}
\thanks{%
The work of Y. Huang and B. D. Rao was supported by Ericsson endowed chair funds, the Center for Wireless Communications, UC Discovery grant com09R-156561 and NSF
grant CCF-1115645. The work of C. W. Tan was supported by grants from the Research Grants Council of Hong Kong Project No. RGC CityU 125212, Qualcomm Inc. and the Science, Technology and Innovation Commission of Shenzhen Municipality, Project No. JCYJ20120829161727318 on Green Communications in Small-cell Mobile Networks. The material in this paper was presented in part at the 46th Annual Conference on Information Sciences and Systems (CISS), Princeton, NJ, March 2012.}
\thanks{%
Y. Huang was with Department of Electrical and Computer Engineering, University
of California, San Diego, La Jolla, CA 92093-0407 USA. He is now with Qualcomm, Corporate R\&D, San Diego, CA, USA (e-mail: yih006@ucsd.edu).}
\thanks{%
C. W. Tan is with Department of Computer Science, City University of Hong Kong,
Kowloon, Hong Kong (e-mail: cheewtan@cityu.edu.hk).}
\thanks{%
B. D. Rao is with Department of Electrical and Computer Engineering,
University of California, San Diego, La Jolla, CA 92093-0407 USA (e-mail: brao@ece.ucsd.edu).}}

\maketitle

%
%

\begin{abstract}
This paper studies joint beamforming and power control in a coordinated multicell downlink system that serves multiple users per cell to
maximize the minimum weighted signal-to-interference-plus-noise ratio. The optimal solution and distributed algorithm with geometrically fast
convergence rate are derived by employing the nonlinear Perron-Frobenius theory and the multicell network duality. The iterative algorithm,
though operating in a distributed manner, still requires instantaneous power update within the coordinated cluster through the backhaul. The
backhaul information exchange and message passing may become prohibitive with increasing number of transmit antennas and increasing number of
users. In order to derive asymptotically optimal solution, random matrix theory is leveraged to design a distributed algorithm that only
requires statistical information. The advantage of our approach is that there is no instantaneous power update through backhaul. Moreover, by
using nonlinear Perron-Frobenius theory and random matrix theory, an effective primal network and an effective dual network are proposed to
characterize and interpret the asymptotic solution.
\end{abstract}


%

\section{Introduction}\label{introduction}
%
%
%
%

\PARstart{T}{o} benefit from the available and increasing spatial degrees of freedom, multicell networks exploit different forms of intercell cooperation to
operate the system in an interference-aware manner \cite{gesbert10, andrews07}. Due to practical constraints such as limited feedback \cite{love08, de12, lozano13} and the
finite capacity of the backhaul \cite{sanderovich09, le11, zhou12}, beamforming level coordination and efficient power control strategies are favored over data level
cooperation and nonlinear precoding approaches \cite{caire03, weingarten06} to effectively scale up the system performance. Considering these
practical constraints, two characteristics are appealing to joint beamforming and power control algorithms design: distributed computation and
fast-convergent algorithms with low complexity. The desired distributed feature addresses system scalability, and the distributed algorithm only
relies on local channel state information (CSI) which can be obtained by uplink measurement in a time division duplex (TDD) system or through
user feedback in a frequency division duplex (FDD) system. On the other hand, simple algorithms possessing fast convergence rate are attractive
in that they reduce the message passing overhead and alleviate the finite backhaul constraint.

The algorithm design is intimately related to the system performance metric of interest. Different system performance metrics reflect different
design priorities. One common approach is to maximize the sum rate of the system. However, due to the non-convexity of the problem, numerically
finding the optimal solution is challenging and the design of distributed algorithms that can compute the global optimal solution efficiently is
still open, e.g., see \cite{karakayali06, christensen08, gesbert10, venturino10, bjornson10, tan11, shi11, bogale12, utschick12, weeraddana13} and the references
therein. It is known that two specific problem formulations admit global optimal solutions: the transmit power minimization subject to
signal-to-interference-plus-noise ratio ($\mathsf{SINR}$) constraints, and the maximization of minimum $\mathsf{SINR}$ subject to power
constraints. The former problem whose priority is energy saving has been addressed extensively in the literature and efficient algorithms have
been proposed for both the single cell and multicell systems \cite{foschini93, yates95, visotsky99, rashid98, bengtsson99, viswanath03, vishwanath03,
schubert04, yu06, song07, dahrouj10, tolli11}. The analysis of single cell downlink relies on the well-known uplink-downlink duality
\cite{rashid98, viswanath03, vishwanath03, schubert04, yu06} which is readily interpreted by the Lagrange duality in convex optimization. In
\cite{song07}, the duality is observed for the MIMO multiuser ad hoc network setting, and in \cite{dahrouj10}, the duality is extended to the
multicell setting.\pubidadjcol

The literature for the latter problem which aims to enforce the fairness level of the system is comparatively less. The max-min $\mathsf{SINR}$
problem was first addressed in \cite{yang98} using an extended coupling matrix approach, and a centralized algorithm was proposed in
\cite{schubert04}, which involves an increased dimension matrix computation. A reformulation of the max-min problem is analyzed in
\cite{wiesel06} by conic programming and a heuristic algorithm is provided. In \cite{huang11}, similar to \cite{wiesel06}, the max-min problem
is tackled from the transmit power minimization perspective and a hierarchical iterative algorithm is proposed. Recently, the problem was
studied in \cite{tan11} using a nonlinear Perron-Frobenius theory \cite{krause01}, and a distributed algorithm was proposed that exhibits the
distributed power control (DPC) structure in \cite{foschini93}. The DPC-like structure is independent of parameter configuration, thus enabling
the application of the power control module in \cite{foschini93} already used in practical cellular systems. The approach \cite{tan11} is
extended to the MIMO downlink in \cite{cai11} wherein the convergence of a heuristic algorithm in \cite{wiesel06} is proved, and a power
optimization problem under multiple power constraint is analyzed in \cite{cai12}. The optimization of the egalitarian fairness i.e., max-min, performance metric also has intimate relationship with other important wireless network performance metric optimization problem, e.g., the weighted sum rate maximization \cite{tan11JSAC, tan11SIAM, zheng13}. The sum rate maximization is nonconvex and NP-hard, and the implication is that fast egalitarian fairness algorithms can be leveraged to solve this nonconvex problem with global optimality guaranteed under special cases. Herein, we firstly extend the analysis in \cite{tan11, cai11} to the multicell setting with multiple serving users per cell. The duality between primal and dual network is derived and characterized by the Perron-Frobenius theory. A distributed algorithm is also proposed which possesses geometrically fast convergence rate.

The designed algorithm, though converging to the optimal solution, requires instantaneous power update within the coordinated cluster through
backhaul. This instantaneous information exchange may become prohibitive when the number of transmit antennas at base station as well as the
serving users per cell grow large. In such emerging large-scale multiple antenna systems \cite{marzetta10, jose11, rusek12, fernandes13}, the
backhaul capability may turn into the bottleneck. In order to alleviate this problem and to enable simplified design that utilizes only the
statistical channel information, additional tools from random matrix theory \cite{tulino04, couillet11} are to be leveraged. The large system
analysis for linear receiver design in the uplink was initiated in \cite{tse99}, and the notion of effective interference and effective
bandwidth was proposed. In \cite{lakshminaryana10}, asymptotic analysis for the transmit power minimization problem is carried out. The
approaches in \cite{huh11} and \cite{wagner11} decouple beamforming and power control by assuming zero-forcing or regularized zero-forcing
beamformers \cite{peel05}. The analysis in \cite{zakhour11} examines the max-min $\mathsf{SINR}$ problem from the transmit power minimization
perspective, and compares several cooperation strategies by assuming a two-cell model with homogeneous channel setting. The analysis is extended
in \cite{zakhour13} for the minimization of the maximum power problem with homogeneous channel setting. In this paper, we perform large system
analysis for the max-min $\mathsf{SINR}$ problem in a general multicell setting. Utilizing tools developed from random matrix theory, the
deterministic equivalents \cite{hachem07, couillet11} for the dual network $\mathsf{SINR}$ and for the primal network $\mathsf{SINR}$ are
established. These asymptotic approximations are used to compute the asymptotic power which only relies on statistical channel information.
Intuitively, in a large-scale multiple antenna system, the optimal powers for different users would approach different deterministic values and
the obtained power can be utilized for optimal beamformer design with local CSI. Moreover, by using nonlinear Perron-Frobenius theory and random
matrix theory, we observe an effective network for the dual network and an effective network for the primal network, which capture the
characteristic of the power control effect in the large system setting. The established effective network is further leveraged to provide a
distributed algorithm with fast convergence rate.

To summarize, the contributions of this paper are three-fold: 1) analysis and algorithm design for joint optimal beamforming and power control
in a finite multicell system to maximize the minimum weighted $\mathsf{SINR}$, 2) the established effective network to characterize the
algebraic structure of the power control problem in the large system setting, and 3) low complexity algorithm design which requires no
instantaneous backhaul exchange. All these contributions lead to efficient methodologies to design algorithms for the large-scale coordinated
multicell downlink. The paper is organized as follows. Section \ref{system} presents the system model. The finite system analysis is provided in
Section \ref{finite}. Section \ref{large} carries out large system analysis and derives the asymptotic solution. Numerical results are presented
in Section \ref{numerical}. Finally, Section \ref{conclusion} concludes the paper.

Notations in this paper are presented as follows. Boldface upper-case letters denote matrices, boldface lower-case letters denote vectors, and
italics denote scalars. The Perron-Frobenius eigenvalue of a nonnegative matrix $\mathbf{F}$ is denoted as $\rho(\mathbf{F})$. Let
$\mathbf{x}(\mathbf{F})$ and $\mathbf{y}(\mathbf{F})$ denote the Perron (right) and left eigenvectors of $\mathbf{F}$ associated with
$\rho(\mathbf{F})$ respectively. $\mathsf{Tr}(\mathbf{A})$ denotes the trace of the matrix $\mathbf{A}$, and $\mathrm{diag}(\mathbf{a})$ denotes
the diagonal matrix having the vector $\mathbf{a}$ on its diagonal. Let $(f(\mathbf{a}))_m$ denote the $m$th element of a function vector
$f(\mathbf{a})$. Let $\mathbf{a}\circ\mathbf{b}\triangleq(a_1b_1,\cdots,a_Mb_M)^{\mathsf{T}}$ (the Schur product). Let $\mathbb{C}$,
$\mathbb{R}_+$, and $\mathbb{R}_{++}$ represent the set of complex numbers, the set of nonnegative real numbers, and the set of positive real
numbers respectively. Let $(\cdot)^{\mathsf{T}}$ and $(\cdot)^{\dag}$ denote the transpose operation and conjugate transpose operation
respectively. $\|\cdot\|$ denotes the Euclidean norm for vectors and spectral norm for matrices, and $\mathop{\longrightarrow}\limits^{a.s.}$
denotes almost sure convergence.

\section{System Model}\label{system}
Consider a coordinated multicell downlink formulated by $J$ coordinating base stations utilizing the same carrier frequency. Each base station
is equipped with $N$ transmit antennas and serves $K$ users simultaneously. Herein, the focus is on the base station side interference
coordination, and each user is assumed to have a single antenna. The received signal $y_{j,k}$ for user $k$ in cell $j$ is written as
\begin{equation}\label{system_eq_1}
y_{j,k}=\sum_{l=1}^{J}\mathbf{h}_{l,j,k}^{\dag}\mathbf{x}_l+z_{j,k}
\end{equation}
where $\mathbf{h}_{l,j,k}\in\mathbb{C}^{N\times 1}$ denotes the channel vector from cell $l$ towards user $k$ in cell $j$,
$\mathbf{x}_l\in\mathbb{C}^{N\times 1}$ is the transmitted signal vector of cell $l$, and $z_{j,k}$ characterizes the additive white noise
effect and any intercell interference not included in the coordinated cluster for user $k$ in cell $j$, which is distributed as
$\mathcal{CN}(0,\sigma_{j,k})$ with $\sigma_{j,k}\in\mathbb{R}_{++}$.

Linear beamforming strategy is assumed at the base station, and thus the transmit signal vector $\mathbf{x}_j$ for cell $j$ can be expressed as
$\mathbf{x}_j=\sum_{k=1}^K \mathbf{x}_{j,k}=\sum_{k=1}^K \sqrt{\frac{p_{j,k}}{N}}s_{j,k}\mathbf{u}_{j,k}$, where
$\mathbf{x}_{j,k}\in\mathbb{C}^{N\times 1}$ represents the signal intended for stream $k$ of cell $j$, $s_{j,k}$ and $\frac{p_{j,k}}{N}$ denote
the information signal and the transmit power for that stream, and $\mathbf{u}_{j,k}\in\mathbb{C}^{N\times 1}$ denotes the normalized transmit
beamformer for user $k$ in cell $j$, i.e., $\|\mathbf{u}_{j,k}\|^2=1$. The $\mathsf{SINR}$ for user $k$ in cell $j$ can be written as
\begin{equation}\label{system_eq_2}
\Gamma_{j,k}^{\mathsf{PN}}\triangleq\mathsf{SINR}_{j,k}^{\mathsf{PN}}=\frac{\frac{p_{j,k}}{N}|\mathbf{h}_{j,j,k}^{\dag}\mathbf{u}_{j,k}|^2}{\mathop{\sum}\limits_{(l,i)\neq(j,k)}\frac{p_{l,i}}{N}|\mathbf{h}_{l,j,k}^{\dag}\mathbf{u}_{l,i}|^2+\sigma_{j,k}}
\end{equation}
where the superscript $(\cdot)^{\mathsf{PN}}$ represents the primal downlink network. Let $w_{j,k}$ denote the weight associated with $p_{j,k}$
for user $k$ in cell $j$ illustrating different power prices, and denote $\beta_{j,k}$ as the priority factor associated with
$\Gamma_{j,k}^{\mathsf{PN}}$ for user $k$ in cell $j$ demonstrating diverse service priorities. Then the max-min problem under weighted sum
power constraint\footnote[1]{The weighted sum power constraint implies that multiple base stations form a coordinated cluster to jointly perform
power control in order to achieve the desired fairness level for users in the cluster.} can be written as follows
\begin{equation}\label{system_eq_3}
\begin{array}{ll}
\mathrm{maximize} & \mathop{\min}\limits_{j,k}\;\frac{\Gamma_{j,k}^{\mathsf{PN}}}{\beta_{j,k}}\\
\mathrm{subject\;to}& \mathop{\sum}\limits_{j,k}w_{j,k}\frac{p_{j,k}}{N}\leq \bar{P},\quad p_{j,k}>0, \quad \|\mathbf{u}_{j,k}\|^2=1\\
\mathrm{variables:} & p_{j,k}, \mathbf{u}_{j,k}.
\end{array}
\end{equation}
The problem (\ref{system_eq_3}) appears non-convex at first, but can be transformed into a second-order cone program \cite{boyd04} by applying
methods similar to that in \cite{wiesel06}, which admits a global optimal solution. However, employing standard convex optimization methods to
find the optimal solution typically requires centralized computation and incurs a fair amount of parameter tuning and message passing overhead
that may not be practical in wireless networks. Thus in Section \ref{finite}, we will employ nonlinear Perron-Frobenius theory to propose
DPC-like algorithm \cite{foschini93} that does not require parameter tuning and has geometrically-fast convergence rate. Then in Section
\ref{large}, algorithms that are even simpler and more practical for systems with a large number of transmit antennas and users\footnote[2]{The large system analysis with algorithm design for a single cell downlink is studied in \cite{huang12c}.} will be
presented by performing an asymptotic analysis.

\section{Finite System Analysis}\label{finite}
This section is devoted to finite system analysis when $N$ and $K$ are not asymptotically large. Section \ref{finite: problem} reformulates
problem (\ref{system_eq_3}) to exploit its analytic structure. Section \ref{finite: duality} establishes the network duality via a
Perron-Frobenius characterization, and provides a geometrically-fast convergent algorithm to compute the optimal solution.

\subsection{Problem Reformulation}\label{finite: problem}
The problem formulation in (\ref{system_eq_3}) essentially regards an interference network with $JK$ users. However, the formulation in terms of
the channel $\mathbf{h}_{l,j,k}$ and the link gain $|\mathbf{h}_{l,j,k}^{\dag}\mathbf{u}_{l,i}|^2$ does not easily lead to amenable analysis. In
order to construct the $JK\times JK$ cross channel interference matrix, consider the matrix $\mathbf{G}\in\mathbb{R}_{++}^{JK\times JK}$ with
subscripts $m$ and $n$, whose entry can be written as
\begin{equation}\label{finite_eq_1}
G_{m,n}=|\mathbf{h}_{\lceil\frac{n}{K}\rceil,\lceil\frac{m}{K}\rceil,m-K\lfloor\frac{m}{K}\rfloor}^{\dag}\mathbf{u}_{\lceil\frac{n}{K}\rceil,n-K\lfloor\frac{n}{K}\rfloor}|^2
\end{equation}
where $\lceil\cdot\rceil$ and $\lfloor\cdot\rfloor$ denote the ceil and floor operation respectively. Thus the channel $\mathbf{h}_{l,j,k}$ can
be represented with subscripts $m$ and $n$:
$\mathbf{h}_{n,m}\triangleq\mathbf{h}_{\lceil\frac{n}{K}\rceil,\lceil\frac{m}{K}\rceil,m-K\lfloor\frac{m}{K}\rfloor}$. Moreover, define the
power vector $\mathbf{p}\in\mathbb{R}_{++}^{JK\times 1}$ as $p_m\triangleq p_{\lceil\frac{m}{K}\rceil,m-K\lfloor\frac{m}{K}\rfloor}$, and the
beamforming matrix as $\mathbb{U}\triangleq(\mathbf{u}_1,\cdots,\mathbf{u}_{JK})$ with
$\mathbf{u}_m\triangleq\mathbf{u}_{\lceil\frac{m}{K}\rceil,m-K\lfloor\frac{m}{K}\rfloor}$. The general formulation in (\ref{finite_eq_1}) can be
easily interpreted through two special cases: a) $J=1$, $K$ arbitrary and b) $K=1$, $J$ arbitrary. The former case refers to a single cell
downlink with $K$ interfering users, while the latter case corresponds to an ad hoc interference network setting with $J$ transmitter-receiver
pairs or a multicell setting with one user served per cell. By the formulation of $\mathbf{G}$, the cross channel interference matrix, denoted
by $\mathbf{F}\in\mathbb{R}_+^{JK\times JK}$ can be obtained by
\begin{equation}\label{finite_eq_2}
F_{m,n}=\left\{
\begin{array}{ll}
0,&\quad \mathrm{if}\;m=n\\ G_{m,n},&\quad \mathrm{if}\;m\neq n.\\
\end{array} \right.
\end{equation}
Similarly, the weight vector $\mathbf{w}\in\mathbb{R}_{++}^{JK\times 1}$, the priority vector $\boldsymbol{\beta}\in\mathbb{R}_{++}^{JK\times
1}$, and the noise vector $\boldsymbol{\sigma}\in\mathbb{R}_{++}^{JK\times 1}$ can be defined by: $w_m\triangleq
w_{\lceil\frac{m}{K}\rceil,m-K\lfloor\frac{m}{K}\rfloor}$, $\beta_m\triangleq\beta_{\lceil\frac{m}{K}\rceil,m-K\lfloor\frac{m}{K}\rfloor}$, and
$\sigma_m\triangleq\sigma_{\lceil\frac{m}{K}\rceil,m-K\lfloor\frac{m}{K}\rfloor}$. From the aforementioned mapping, if we denote the
$\mathsf{SINR}$ vector as $\boldsymbol{\Gamma}^{\mathsf{PN}}\in\mathbb{R}_{++}^{JK\times 1}$ with
$\Gamma_m^{\mathsf{PN}}\triangleq\Gamma_{\lceil\frac{m}{K}\rceil,m-K\lfloor\frac{m}{K}\rfloor}^{\mathsf{PN}}$, and the auxiliary vector
$\mathbf{g}\in\mathbb{R}_{++}^{JK\times 1}$ with $\mathbf{g}\triangleq\left(\frac{1}{G_{1,1}},\cdots,\frac{1}{G_{JK,JK}}\right)^{\mathsf{T}}$,
then the optimization problem (\ref{system_eq_3}) can be readily reformulated as follows:
\begin{equation}\label{finite_eq_3}
\begin{array}{ll}
\mathrm{maximize} & \mathop{\min}\limits_m\;\frac{\Gamma_m^{\mathsf{PN}}(\mathbf{p},\mathbb{U})}{\beta_m}=\frac{\frac{1}{N}p_m}{\left(\mathrm{diag}(\boldsymbol{\beta}\circ\mathbf{g})\left(\frac{1}{N}\mathbf{Fp}+\boldsymbol{\sigma}\right)\right)_m}\\
\mathrm{subject\;to}& \frac{1}{N}\mathbf{w}^{\mathsf{T}}\mathbf{p}\leq \bar{P},\quad \mathbf{p}>0, \quad \|\mathbf{u}_m\|^2=1\\
\mathrm{variables:} & \mathbf{p}, \mathbb{U}.
\end{array}
\end{equation}
It can be shown that solving (\ref{finite_eq_3}) is equivalent to solving (\ref{system_eq_3}). The compact formulation in (\ref{finite_eq_3})
introduces a nonnegative matrix
$\mathrm{diag}(\boldsymbol{\beta}\circ\mathbf{g})\left(\mathbf{F}+(1/\bar{P})\boldsymbol{\sigma}\mathbf{w}^{\mathsf{T}}\right)$, whose algebraic
structure helps in establishing the network duality and is pursued next.

\subsection{Network Duality and Algorithm Design}\label{finite: duality}
The analytic structure in (\ref{finite_eq_3}) is similar to the formulation in \cite{cai11} for the single cell multiuser downlink scenario. In
\cite{cai11}, the uplink-downlink duality is proved by a geometric programming formulation and the Lagrange duality. Herein, we provide a
network duality interpretation for the max-min based multicell scenario via Perron-Frobenius characterization.

For any given beamforming matrix $\mathbb{U}$, a simpler optimization problem for (\ref{finite_eq_3}) can be formulated by only optimizing the
power solution. It is known that at optimality, the weighted $\mathsf{SINR}$ for different users are the same, and the weighted power constraint
becomes tight \cite{tan11}. Now if we explicitly make the dependence on $\mathbb{U}$ and denote the optimal weighted $\mathsf{SINR}$ as
$\tau^*(\mathbb{U})$, then the optimal power solution satisfies \cite{tan11, cai11}:
\begin{equation}\label{finite_eq_4}
\frac{1}{\tau^*(\mathbb{U})}\frac{\mathbf{p}^*(\mathbb{U})}{N}=\mathrm{diag}(\boldsymbol{\beta}\circ\mathbf{g}(\mathbb{U}))\left(\mathbf{F}(\mathbb{U})+(1/\bar{P})\boldsymbol{\sigma}\mathbf{w}^{\mathsf{T}}\right)\frac{\mathbf{p}^*(\mathbb{U})}{N}.
\end{equation}
From (\ref{finite_eq_4}), it can be shown from nonnegative matrix theory \cite{berman79} that $\frac{\mathbf{p}^*(\mathbb{U})}{N}$ is the Perron
(right) eigenvector (up to a scaling factor) of the nonnegative matrix
$\mathrm{diag}(\boldsymbol{\beta}\circ\mathbf{g}(\mathbb{U}))\left(\mathbf{F}(\mathbb{U})+(1/\bar{P})\boldsymbol{\sigma}\mathbf{w}^{\mathsf{T}}\right)$,
namely,
$\frac{\mathbf{p}^*(\mathbb{U})}{N}=\frac{N\bar{P}\mathbf{x}\left(\mathrm{diag}(\boldsymbol{\beta}\circ\mathbf{g}(\mathbb{U}))\left(\mathbf{F}(\mathbb{U})+(1/\bar{P})\boldsymbol{\sigma}\mathbf{w}^{\mathsf{T}}\right)\right)}{\mathbf{w}^{\mathsf{T}}\mathbf{x}\left(\mathrm{diag}(\boldsymbol{\beta}\circ\mathbf{g}(\mathbb{U}))\left(\mathbf{F}(\mathbb{U})+(1/\bar{P})\boldsymbol{\sigma}\mathbf{w}^{\mathsf{T}}\right)\right)}$,
and $\tau^*(\mathbb{U})$ is related to its spectral radius by the following:
\begin{equation}\label{finite_eq_5}
\tau^*(\mathbb{U})=\frac{1}{\rho\left(\mathrm{diag}(\boldsymbol{\beta}\circ\mathbf{g}(\mathbb{U}))\left(\mathbf{F}(\mathbb{U})+(1/\bar{P})\boldsymbol{\sigma}\mathbf{w}^{\mathsf{T}}\right)\right)}.
\end{equation}

Now, in order to establish the network duality, consider the hypothesized dual uplink network by construction. Denote the dual network transmit
power solution $\mathbf{q}\in\mathbb{R}_{++}^{JK\times 1}$ as $q_m\triangleq q_{\lceil\frac{m}{K}\rceil,m-K\lfloor\frac{m}{K}\rfloor}$, where
$\frac{q_{j,k}}{N}$ with subscripts $j$ and $k$ denotes the reciprocal uplink transmit power for user $k$ in cell $j$. Let the weight vector
$\mathbf{w}$ in the primal network be the noise vector in the dual network, and conversely let the noise vector $\boldsymbol{\sigma}$ in the
primal network be the weight vector in the dual network. Then the max-min problem for the dual network given receive beamforming matrix
$\mathbb{U}$ can be formulated as

\vspace{-10pt}
{\small
\begin{equation}\label{finite_eq_6}
\begin{array}{ll}
\mathrm{maximize} & \mathop{\min}\limits_m\;\frac{\Gamma_m^{\mathsf{DN}}(\mathbf{q},\mathbb{U})}{\beta_m}=\frac{\frac{1}{N}q_m(\mathbb{U})}{\left(\mathrm{diag}(\boldsymbol{\beta}\circ\mathbf{g}(\mathbb{U}))\left(\frac{1}{N}\mathbf{F}^{\mathsf{T}}(\mathbb{U})\mathbf{q}(\mathbb{U})+\mathbf{w}\right)\right)_m}\\
\mathrm{subject\;to}& \frac{1}{N}\boldsymbol{\sigma}^{\mathsf{T}}\mathbf{q}(\mathbb{U})\leq \bar{P},\quad \mathbf{q}(\mathbb{U})>0\\
\mathrm{variables:} & \mathbf{q}(\mathbb{U})
\end{array}
\end{equation}
}%
where the superscript $(\cdot)^{\mathsf{DN}}$ denotes the dual uplink network. By leveraging the following properties of nonnegative matrices
\cite{berman79}: $\rho(\mathbf{A})=\rho(\mathbf{A}^{\mathsf{T}})$ and $\rho(\mathbf{AB})=\rho(\mathbf{BA})$, the optimal solution for
(\ref{finite_eq_6}) equals
$\frac{1}{\rho\left(\mathrm{diag}(\boldsymbol{\beta}\circ\mathbf{g}(\mathbb{U}))\left(\mathbf{F}^{\mathsf{T}}(\mathbb{U})+(1/\bar{P})\mathbf{w}\boldsymbol{\sigma}^{\mathsf{T}}\right)\right)}$.
Comparing with the optimal solution for the primal network in (\ref{finite_eq_5}), the network duality is observed by employing
$\mathbf{F}^{\mathsf{T}}$ as the cross channel interference matrix for the dual network and reversing the role of $\mathbf{w}$ and
$\boldsymbol{\sigma}$. Thus the optimal power solution given $\mathbb{U}$ is the left eigenvector of the nonnegative matrix
$\mathrm{diag}(\boldsymbol{\beta}\circ\mathbf{g}(\mathbb{U}))\left(\mathbf{F}(\mathbb{U})+(1/\bar{P})\boldsymbol{\sigma}\mathbf{w}^{\mathsf{T}}\right)$,
namely,
$\frac{\mathbf{q}^*(\mathbb{U})}{N}=\frac{N\bar{P}\mathbf{y}\left(\mathrm{diag}(\boldsymbol{\beta}\circ\mathbf{g}(\mathbb{U}))\left(\mathbf{F}(\mathbb{U})+(1/\bar{P})\boldsymbol{\sigma}\mathbf{w}^{\mathsf{T}}\right)\right)}{\boldsymbol{\sigma}^{\mathsf{T}}\mathbf{y}\left(\mathrm{diag}(\boldsymbol{\beta}\circ\mathbf{g}(\mathbb{U}))\left(\mathbf{F}(\mathbb{U})+(1/\bar{P})\boldsymbol{\sigma}\mathbf{w}^{\mathsf{T}}\right)\right)}$.
Note that since the network duality holds for any given $\mathbb{U}$, the achievable $\mathsf{SINR}$ regions of the max-min problem are the same
for both the primal network and the dual network.

\begin{table}[t]
\caption{Algorithm A: Max-min weighted $\mathsf{SINR}$ for multicell downlink}\label{tab_1} \centering \rule{\linewidth}{0.1mm}
\begin{itemize}
\item Initialize arbitrary $\mathbf{p}[0]\in\mathbb{R}_{++}^{JK\times 1}$, $\mathbf{q}[0]\in\mathbb{R}_{++}^{JK\times 1}$ and $\mathbf{u}_m[0]\in\mathbb{C}^{N\times 1}$ for
$m=1,\ldots,JK$ such that $\|\mathbf{u}_m[0]\|=1,\forall m$, $\frac{1}{N}\mathbf{w}^{\mathsf{T}}\mathbf{p}[0]\leq\bar{P}$, and
$\frac{1}{N}\boldsymbol{\sigma}^{\mathsf{T}}\mathbf{q}[0]\leq\bar{P}$.
\end{itemize}
\begin{enumerate}
\item Update dual network power $\mathbf{q}[\kappa+1]$:
\begin{equation*}
q_m[\kappa+1]=\left(\frac{\beta_m}{\Gamma_m^{\mathsf{DN}}(\mathbf{q}[\kappa],\mathbb{U}[\kappa])}\right)q_m[\kappa] \quad \forall m.
\end{equation*}
\item Normalize $\mathbf{q}[\kappa+1]$:
\begin{equation*}
\mathbf{q}[\kappa+1]\leftarrow\frac{N\bar{P}}{\boldsymbol{\sigma}^{\mathsf{T}}\mathbf{q}[\kappa+1]}\mathbf{q}[\kappa+1].
\end{equation*}
\item Update transmit beamforming matrix $\mathbb{U}[\kappa+1]$:
\begin{equation*}
\mathbf{u}_m[\kappa+1]=\frac{(\sum_{n\neq
m}\frac{q_n[\kappa+1]}{N}\mathbf{h}_{m,n}\mathbf{h}_{m,n}^{\dag}+w_m\mathbf{I})^{-1}\mathbf{h}_{m,m}}{\|(\sum_{n\neq
m}\frac{q_n[\kappa+1]}{N}\mathbf{h}_{m,n}\mathbf{h}_{m,n}^{\dag}+w_m\mathbf{I})^{-1}\mathbf{h}_{m,m}\|} \quad \forall m.
\end{equation*}
\item Update primal network power $\mathbf{p}[\kappa+1]$:
\begin{equation*}
p_m[\kappa+1]=\left(\frac{\beta_m}{\Gamma_m^{\mathsf{PN}}(\mathbf{p}[\kappa],\mathbb{U}[\kappa+1])}\right)p_m[\kappa] \quad \forall m.
\end{equation*}
\item Normalize $\mathbf{p}[\kappa+1]$:
\begin{equation*}
\mathbf{p}[\kappa+1]\leftarrow\frac{N\bar{P}}{\mathbf{w}^{\mathsf{T}}\mathbf{p}[\kappa+1]}\mathbf{p}[\kappa+1].
\end{equation*}
\end{enumerate}
\rule{\linewidth}{0.1mm}

\end{table}

The motivation for establishing the dual network is to exploit the decoupled property of the receive beamformer optimization and to utilize the
optimized received beamformer as the optimal transmit beamformer for each user. The optimal beamforming matrix $\mathbb{U}^*$ depends on the
power vector $\mathbf{q}$, and for any given $\mathbf{q}$, the optimal beamformer $\mathbf{u}_m^*(\mathbf{q})$ can be obtained by
\begin{equation}\label{finite_eq_7}
\mathbf{u}_m^*(\mathbf{q})=\arg\min_{\mathbf{u}_m}\frac{\mathbf{u}_m^{\dag}(\sum_{n\neq
m}\frac{q_n}{N}\mathbf{h}_{m,n}\mathbf{h}_{m,n}^{\dag}+w_m\mathbf{I})\mathbf{u}_m}{\mathbf{u}_m^{\dag}\mathbf{h}_{m,m}\mathbf{h}_{m,m}^{\dag}\mathbf{u}_m}
\end{equation}
which can be readily solved and is known to be the minimum variance distortionless response (MVDR) beamformer which is given by:
\begin{equation}\label{finite_eq_8}
\mathbf{u}_m^*(\mathbf{q})=\frac{(\sum_{n\neq
m}\frac{q_n}{N}\mathbf{h}_{m,n}\mathbf{h}_{m,n}^{\dag}+w_m\mathbf{I})^{-1}\mathbf{h}_{m,m}}{\|(\sum_{n\neq
m}\frac{q_n}{N}\mathbf{h}_{m,n}\mathbf{h}_{m,n}^{\dag}+w_m\mathbf{I})^{-1}\mathbf{h}_{m,m}\|}.
\end{equation}
Therefore, the optimal solution for the beamformers, the power of the dual network, and the power of the primal network can be written as:
$\mathbf{u}_m^*=\mathbf{u}_m^*(\mathbf{q}^*)$, $\mathbf{q}^*=\mathbf{q}^*(\mathbb{U}^*)$, and $\mathbf{p}^*=\mathbf{p}^*(\mathbb{U}^*)$. The
optimal solution is of analytical interest. In order to derive a fast algorithm to compute the optimal solution in a distributed manner, we
employ nonlinear Perron-Frobenius theory and propose the algorithm given in Table \ref{tab_1}, referred to as \textit{Algorithm A} for the
multicell scenario. It exhibits the DPC-like structure as in \cite{tan11, cai11} for the single cell scenario. The convergence property of
\textit{Algorithm A} is discussed in the following theorem.
\begin{theorem}\label{theorem_1}
Starting from any initial point $\mathbf{q}[0]$, $\mathbf{p}[0]$, and $\mathbb{U}[0]$, the $\mathbf{q}[\kappa]$, $\mathbf{p}[\kappa]$, and
$\mathbb{U}[\kappa]$ in \textit{Algorithm A} converges geometrically fast to the optimal solution $\mathbf{q}^*$, $\mathbf{p}^*$, and
$\mathbb{U}^*$.
\end{theorem}
\begin{proof}
The proof is given in Appendix A.
\end{proof}
\textit{Remark:} Distributed algorithms utilizing only local CSI and requiring limited backhaul exchange are important for practical
implementation issues. \textit{Algorithm A} is distributed in the sense that the iterative update (step $1$, $3$, $4$ of \textit{Algorithm A})
can be independently performed for each individual user at each base station. In addition, each base station only employs local CSI, which can
be directly obtained in a TDD system or acquired by user feedback in a FDD system. The normalization procedure (step $2$ and $5$ of of
\textit{Algorithm A}), however, requires a central computation of $\mathbf{w}^{\mathsf{T}}\mathbf{p}[\kappa]$ and
$\boldsymbol{\sigma}^{\mathsf{T}}\mathbf{q}[\kappa]$. This procedure can be made distributed by gossip algorithms \cite{boyd06} and power update
through the backhaul.

Hitherto, an algorithm for computing the optimal solution to (\ref{system_eq_3}) is established. In Section \ref{numerical}, we provide
numerical results that support and confirm its fast convergence property. Furthermore, with minimal parameter exchange and configuration, this
algorithm is practical in a finite system. However, in a large-scale system when both $N$ and $K$ become large, the instantaneous power update
across the coordinated cluster limits its practical implementation. Therefore, a lower complexity algorithm is needed in large-scale systems and
is studied next in Section \ref{large}.

\section{Large System Analysis}\label{large}
This section is devoted to a large system analysis when both the number of transmit antennas $N$ and the number of serving users per cell $K$ go to infinity while the ratio (load factor) $\lim\frac{K}{N}$ remains bounded, i.e., the notation $N\rightarrow\infty$ denotes that both $N$ and
$K$ become large, while $\lim\inf\frac{K}{N}>0$ and $\lim\sup\frac{K}{N}<\infty$. In this large-scale system setting, for a given channel
realization, the amount of instantaneous power update through the limited backhaul can be impractically large and thus impacts the system
performance. A key question is whether it is possible to design a non-iterative algorithm to compute the beamformer and still achieve some form
of optimal egalitarian fairness. Herein, the optimality is in the asymptotic sense. This means that, if the power $\mathbf{p}$ and $\mathbf{q}$
in the large system converge to some deterministic values that only rely on statistical channel information, then these deterministic values can
be a priori calculated, stored, and updated only when the channel statistics change\footnote[3]{This idea relates to algorithmic developments in
the context of and in support of the design of situational aware wireless networks \cite{huang12Thesis}. The envisioned situational aware wireless networks adapt system parameters and
algorithms design to the channel attributes (i.e., different types of channel information, various
channel statistics reflected in different dimensions), user attributes (i.e., different user densities, fairness requirements, user mobilities), and system attributes (i.e., backhaul capabilities, large system or sparse system structures, energy efficiencies),
which constitute the wireless environment and network situations, see \cite{huangJ1, huangJ4, yarkan08, ghosh10, huang12Globecom, li13, cho13, huangJ5, hong13, hu10, aliu13} for examples driving this trend.}. Thereafter, the beamforming matrix ought to be
non-iteratively computed using these slowly updated power values and the available instantaneous local CSI.

This idea of practical implementation for large systems will be studied by addressing two problems related to (\ref{system_eq_3}). Firstly,
different users in the multicell network have potentially different weights, different priorities, different noise powers, and more importantly,
different large-scale channel effects which may consist of path loss, shadowing, and antenna gain. Thus, to maintain the max-min fairness across
users, the powers for different users would converge to different deterministic values in the large system setting. One key issue is to
establish the asymptotic optimality for both the dual network power $\mathbf{q}$ and primal network power $\mathbf{p}$. Another key issue is to
design distributed algorithm to compute these deterministic values.

In Section \ref{finite}, no specific channel models are assumed. Now for amenable analysis, the transformed notation using subscripts $m$ and
$n$ will be still employed and the following channel model is further assumed:
\begin{equation}\label{large_eq_1}
\mathbf{h}_{m,n}=\sqrt{d_{m,n}}\tilde{\mathbf{h}}_{m,n}
\end{equation}
where $d_{m,n}$ represents the large-scale channel effect and illustrates the statistical channel information. The $\tilde{\mathbf{h}}_{m,n}$
denotes the normalized CSI whose elements are independent and identically distributed as $\mathcal{CN}(0,1)$. This assumption corresponds to the
practical setting where the antenna elements equipped at each base station are placed sufficiently apart. Herein, independent channel assumption
is employed and the analysis with the general correlated channel model \cite{dumont10, taricco11, artigue11, couillet11j, zhang12} is left for
future work. Employing this channel model, the asymptotic analysis for the dual network and primal network is carried out in Section \ref{large:
dual} and Section \ref{large: primal}, respectively.

\subsection{Asymptotic Analysis for the Dual Network}\label{large: dual}
The large system analysis for the dual network is examined first to derive the asymptotic dual network power, which is utilized for beamformer
design. One key step is to study the asymptotic behavior of the dual network $\mathsf{SINR}$, whose expression is given by using the optimal
MVDR beamformer as follows:
{\small
\begin{align}
&\Gamma_m^{\mathsf{DN}}(\mathbf{q})=\frac{q_m}{N}\mathbf{h}_{m,m}^{\dag}\left(\sum_{n\neq
m}\frac{q_n}{N}\mathbf{h}_{m,n}\mathbf{h}_{m,n}^{\dag}+w_m\mathbf{I}\right)^{-1}\mathbf{h}_{m,m}\notag\\
\label{large_eq_2} &=\frac{q_md_{m,m}}{N}\tilde{\mathbf{h}}_{m,m}^{\dag}\left(\sum_{n\neq
m}\frac{q_nd_{m,n}}{N}\tilde{\mathbf{h}}_{m,n}\tilde{\mathbf{h}}_{m,n}^{\dag}+w_m\mathbf{I}\right)^{-1}\tilde{\mathbf{h}}_{m,m}\;\forall m.
\end{align}
}

Since each instantaneous CSI is random, the instantaneous $\mathsf{SINR}$ in (\ref{large_eq_2}) is a random variable in quadratic form.
Moreover, since the dual network power and large scale channel effects are diverse across users, if we define the random matrix
$\tilde{\mathbf{H}}_m$ as $\tilde{\mathbf{H}}_m\tilde{\mathbf{H}}_m^{\dag}\triangleq\sum_{n\neq
m}\frac{q_nd_{m,n}}{N}\tilde{\mathbf{h}}_{m,n}\tilde{\mathbf{h}}_{m,n}^{\dag}$, then the random matrix $\tilde{\mathbf{H}}_m$ possesses a
variance profile \cite{hachem07, kammoun09}. The asymptotic approximation for $\Gamma_m^{\mathsf{DN}}(\mathbf{q})$ is given in the following
lemma.
\begin{lemma}\label{lemma_1}
The instantaneous random variable $\Gamma_m^{\mathsf{DN}}(\mathbf{q})$ can be approximated by a deterministic quantity\footnote[4]{Note that we
present the asymptotic behavior of $\Gamma_m^{\mathsf{DN}}(\mathbf{q})$ with a given power vector $\mathbf{q}$, not with the instantaneous
optimal power vector $\mathbf{q}^*$. The instantaneous optimal power vector is a function of channel and thus complicates standard large system
analysis. Bounding techniques trying to investigate this issue are conducted in \cite{zakhour11}. In this paper, iterative method is used to
compute the asymptotically optimal power. This comment carries over to the following lemmas.} $\gamma_m^{\mathsf{DN}}(\mathbf{q})$ such that
$\Gamma_m^{\mathsf{DN}}(\mathbf{q})-\gamma_m^{\mathsf{DN}}(\mathbf{q})\mathop{\longrightarrow}\limits^{a.s.}0$ as the system dimension
$N\rightarrow\infty$. Also, $\gamma_m^{\mathsf{DN}}(\mathbf{q})$ is described by the following fixed-point equation:
\begin{equation}\label{large_eq_3}
\gamma_m^{\mathsf{DN}}(\mathbf{q})=\frac{q_md_{m,m}}{w_m+\frac{1}{N}\mathop{\sum}\limits_{n\neq
m}\frac{q_nq_md_{m,n}d_{m,m}}{q_md_{m,m}+q_nd_{m,n}\gamma_m^{\mathsf{DN}}(\mathbf{q})}} \quad \forall m.
\end{equation}
\end{lemma}
\begin{proof}
The proof is given in Appendix B.
\end{proof}
From Lemma \ref{lemma_1}, we know that $\gamma_m^{\mathsf{DN}}(\mathbf{q})$ becomes more accurate when increasing the system dimension, and is
asymptotically tight for $\Gamma_m^{\mathsf{DN}}(\mathbf{q})$. For further analysis, an auxiliary vector
$\boldsymbol{\phi}\in\mathbb{R}_{++}^{JK\times 1}$ is defined with $\phi_m(\mathbf{q})\triangleq
\frac{\gamma_m^{\mathsf{DN}}(\mathbf{q})}{q_md_{m,m}},\;\forall m$. Then from Lemma \ref{lemma_1}, the fixed-point equation for
$\phi_m(\mathbf{q})$ can be written as
\begin{equation}\label{large_eq_4}
\phi_m(\mathbf{q})=\frac{1}{w_m+\frac{1}{N}\mathop{\sum}\limits_{n\neq m}\frac{q_nd_{m,n}}{1+q_nd_{m,n}\phi_m(\mathbf{q})}} \quad \forall m.
\end{equation}
From (\ref{large_eq_4}), it is easy to see that $\mathbf{q}$ and $\boldsymbol{\phi}$ are coupled and their relationship only depends on the
statistical channel information reflected in $d_{m,n}$. Designing algorithms to compute $\mathbf{q}$ and $\boldsymbol{\phi}(\mathbf{q})$ is of
primary interest and one common approach is to examine the conditional convergence property of $\mathbf{q}$ and $\boldsymbol{\phi}$ separately.

\begin{table}[t]
\caption{Algorithm B: Computation of $\hat{\boldsymbol{\phi}}$ given $\hat{\mathbf{q}}$}\label{tab_2} \centering \rule{\linewidth}{0.1mm}
\begin{itemize}
\item Initialize arbitrary $\hat{\boldsymbol{\phi}}[0]\in\mathbb{R}_{++}^{JK\times 1}$ with a given $\hat{\mathbf{q}}$.
\item Update $\hat{\boldsymbol{\phi}}[\ell+1]$:
\begin{equation*}
\hat{\phi}_m[\ell+1]=\frac{1}{w_m+\frac{1}{N}\mathop{\sum}\limits_{n\neq m}\frac{\hat{q}_nd_{m,n}}{1+\hat{q}_nd_{m,n}\hat{\phi}_m[\ell]}} \quad
\forall m.
\end{equation*}
\end{itemize}
\rule{\linewidth}{0.1mm}

\end{table}

The convergence property of $\boldsymbol{\phi}$ given $\mathbf{q}$ is relatively easy to establish since it does not involve any constraint.
Given any $\hat{\mathbf{q}}$ satisfying the dual network power constraint, the algorithm to compute the corresponding
$\hat{\boldsymbol{\phi}}(\hat{\mathbf{q}})$ is given in Table \ref{tab_2} and is referred to as $\textit{Algorithm B}$ whose convergence
property is given below.
\begin{lemma}\label{lemma_2}
For a given $\hat{\mathbf{q}}$, starting from any initial $\hat{\boldsymbol{\phi}}[0]$, the $\hat{\boldsymbol{\phi}}[\ell]$ in
$\textit{Algorithm B}$ converges to the unique solution\footnote[5]{The existence of the solution can be shown by employing the same method as
in the proof of Theorem $6.1$ in \cite{couillet11} and the proof of Theorem $1$ in \cite{couillet11j}.} of the fixed-point equation
(\ref{large_eq_4}).
\end{lemma}
\begin{proof}
The proof is given in Appendix B.
\end{proof}
Now consider the convergence property of $\mathbf{q}$ given $\boldsymbol{\phi}$. Combining (\ref{large_eq_3}) and (\ref{large_eq_4}) yields the
equivalent fixed-point equation for $\gamma_m^{\mathsf{DN}}(\mathbf{q})=\frac{q_md_{m,m}}{w_m+\frac{1}{N}\mathop{\sum}\limits_{n\neq
m}\frac{q_nd_{m,n}}{1+q_nd_{m,n}\phi_m(\mathbf{q})}}$. Thus the additive effect of $\frac{1}{N}\mathop{\sum}\limits_{n\neq
m}\frac{q_nd_{m,n}}{1+q_nd_{m,n}\phi_m(\mathbf{q})}$ can be seen as the asymptotically equivalent interference and is regarded as effective
interference in \cite{tse99}. In the sequel, we construct the \textit{effective dual network} to draw further insight for the power control
problem.

Firstly, the following power control problem conditioned on $\hat{\boldsymbol{\phi}}$ is constructed by considering the weighted power
constraint:
\begin{equation}\label{large_eq_5}
\begin{array}{ll}
\mathrm{maximize} & \mathop{\min}\limits_m\;\frac{\hat{q}_md_{m,m}}{\beta_m\left(w_m+\frac{1}{N}\mathop{\sum}\limits_{n\neq
m}\frac{\hat{q}_nd_{m,n}}{1+\hat{q}_nd_{m,n}\hat{\phi}_m}\right)}\\
\mathrm{subject\;to}& \frac{1}{N}\boldsymbol{\sigma}^{\mathsf{T}}\hat{\mathbf{q}}\leq \bar{P},\quad \hat{\mathbf{q}}>0\\
\mathrm{variables:} & \hat{\mathbf{q}}.
\end{array}
\end{equation}
Then, by defining the vector $\mathbf{e}^{\mathsf{DN}}\triangleq\left(\frac{1}{d_{1,1}},\cdots,\frac{1}{d_{JK,JK}}\right)^{\mathsf{T}}$ and the
nonnegative matrix $\mathbf{E}^{\mathsf{DN}}(\hat{\mathbf{q}})$ as
\begin{equation}\label{large_eq_6}
E_{m,n}^{\mathsf{DN}}(\hat{\mathbf{q}})=\left\{
\begin{array}{ll}
0,&\quad \mathrm{if}\;m=n\\ \frac{d_{m,n}}{1+\hat{q}_nd_{m,n}\hat{\phi}_m},&\quad \mathrm{if}\;m\neq n\\
\end{array} \right.
\end{equation}
the objective function in (\ref{large_eq_5}) can be expressed compactly as
$\frac{\hat{q}_m}{\left(\mathrm{diag}(\boldsymbol{\beta}\circ\mathbf{e}^{\mathsf{DN}})\left(\frac{1}{N}\mathbf{E}^{\mathsf{DN}}(\hat{\mathbf{q}})\hat{\mathbf{q}}+\mathbf{w}\right)\right)_m}$,
whose algebraic structure leads to the following eigenvalue problem in terms of the power $\hat{\mathbf{q}}^*$ and weighted asymptotic
$\mathsf{SINR}$ $\varsigma^*$:
\begin{equation}\label{large_eq_7}
\frac{\hat{\mathbf{q}}^*}{\varsigma^*}=\mathrm{diag}\left(\boldsymbol{\beta}\circ\mathbf{e}^{\mathsf{DN}}\right)\left(\mathbf{E}^{\mathsf{DN}}(\hat{\mathbf{q}}^*)+(1/\bar{P})\mathbf{w}\boldsymbol{\sigma}^{\mathsf{T}}\right)\frac{\hat{\mathbf{q}}^*}{N}.
\end{equation}
By comparing with (\ref{finite_eq_4}), we can see that $\mathbf{E}^{\mathsf{DN}}(\hat{\mathbf{q}})$ can be regarded as the effective cross
channel interference matrix and the effective dual network can be characterized by the nonnegative matrix
$\mathrm{diag}\left(\boldsymbol{\beta}\circ\mathbf{e}^{\mathsf{DN}}\right)\left(\mathbf{E}^{\mathsf{DN}}(\hat{\mathbf{q}})+(1/\bar{P})\mathbf{w}\boldsymbol{\sigma}^{\mathsf{T}}\right)$.
Note that in the finite system setting, the cross channel interference matrix $\mathbf{F}$ is independent of the power. However, for the large
system setting, $\mathbf{E}^{\mathsf{DN}}(\hat{\mathbf{q}})$ and $\hat{\mathbf{q}}$ are interdependent. In the following, we employ nonlinear
Perron-Frobenius theory to propose a distributed algorithm to compute $\hat{\mathbf{q}}^*$ given $\hat{\boldsymbol{\phi}}$, which is given in
Table \ref{tab_3} and is referred to as \textit{Algorithm C}.

\begin{table}[t]
\caption{Algorithm C: Computation of $\hat{\mathbf{q}}$ given $\hat{\boldsymbol{\phi}}$}\label{tab_3} \centering \rule{\linewidth}{0.1mm}
\begin{itemize}
\item Initialize arbitrary $\hat{\mathbf{q}}[0]\in\mathbb{R}_{++}^{JK\times 1}$ with a given $\hat{\boldsymbol{\phi}}$ such that $\frac{1}{N}\boldsymbol{\sigma}^{\mathsf{T}}\hat{\mathbf{q}}[0]\leq\bar{P}$.
\end{itemize}
\begin{enumerate}
\item Update dual network power $\hat{\mathbf{q}}[\ell+1]$:
\begin{equation*}
\hat{q}_m[\ell+1]=\frac{\beta_m}{d_{m,m}}\left(w_m+\frac{1}{N}\sum_{n\neq
m}\frac{\hat{q}_n[\ell]d_{m,n}}{1+\hat{q}_n[\ell]d_{m,n}\hat{\phi}_m}\right) \quad \forall m.
\end{equation*}
\item Normalize $\hat{\mathbf{q}}[\ell+1]$:
\begin{equation*}
\hat{\mathbf{q}}[\ell+1]\leftarrow\frac{N\bar{P}}{\boldsymbol{\sigma}^{\mathsf{T}}\hat{\mathbf{q}}[\ell+1]}\hat{\mathbf{q}}[\ell+1].
\end{equation*}
\end{enumerate}
\rule{\linewidth}{0.1mm}

\end{table}

\begin{theorem}\label{theorem_2}
For a given $\hat{\boldsymbol{\phi}}$, starting from any initial $\hat{\mathbf{q}}[0]$, the $\hat{\mathbf{q}}[\ell]$ in $\textit{Algorithm C}$
converges geometrically fast to the optimal solution $\hat{\mathbf{q}}^*(\hat{\boldsymbol{\phi}})$ of (\ref{large_eq_5}).
\end{theorem}
\begin{proof}
The proof is given in Appendix A.
\end{proof}
After establishing the convergence properties of computing $\hat{\boldsymbol{\phi}}(\hat{\mathbf{q}})$ in \textit{Algorithm B} and
$\hat{\mathbf{q}}^*(\hat{\boldsymbol{\phi}})$ in \textit{Algorithm C}, both \textit{Algorithm B} and \textit{Algorithm C} can be combined using
alternate optimization to compute an asymptotically local optimal solution $\boldsymbol{\phi}(\hat{\mathbf{q}}^*)$ and $\hat{\mathbf{q}}^*$. The
asymptotically optimal dual network power is used to design the asymptotically optimal beamformer in (\ref{finite_eq_8}). The procedure to
derive the asymptotically optimal primal network power requires $\hat{\mathbf{q}}^*$ and $\boldsymbol{\phi}(\hat{\mathbf{q}}^*)$, and is pursued
next.

\subsection{Asymptotic Analysis for the Primal Network}\label{large: primal}
Similar procedure for analyzing the dual network can be applied to the primal network in order to examine the asymptotically optimal transmit
power $\hat{\mathbf{p}}^*$. From the analysis in Section \ref{finite}, the primal network $\mathsf{SINR}$ is given as
\begin{equation}\label{large_eq_8}
\Gamma_m^{\mathsf{PN}}(\mathbf{p})=\frac{\frac{p_md_{m,m}}{N}|\tilde{\mathbf{h}}_{m,m}^{\dag}\mathbf{u}_m^*|^2}{\mathop{\sum}\limits_{n\neq
m}\frac{p_nd_{n,m}}{N}|\tilde{\mathbf{h}}_{n,m}^{\dag}\mathbf{u}_n^*|^2+\sigma_m}.
\end{equation}
The asymptotic approximation of $\Gamma_m^{\mathsf{PN}}(\mathbf{p})$ is presented in the following lemma.
\begin{lemma}\label{lemma_3}
The instantaneous random variable $\Gamma_m^{\mathsf{PN}}(\mathbf{p})$ can be approximated by a deterministic quantity
$\gamma_m^{\mathsf{PN}}(\mathbf{p})$ such that
$\Gamma_m^{\mathsf{PN}}(\mathbf{p})-\gamma_m^{\mathsf{PN}}(\mathbf{p})\mathop{\longrightarrow}\limits^{a.s.}0$ as the system dimension
$N\rightarrow\infty$. Also, $\gamma_m^{\mathsf{PN}}(\mathbf{p})$ is described by the following equation:
\begin{equation}\label{large_eq_9}
\gamma_m^{\mathsf{PN}}(\mathbf{p})=\frac{p_md_{m,m}\frac{\phi_m^2(\mathbf{q})}{-\phi_m'(\mathbf{q})}}{\sigma_m+\frac{1}{N}\mathop{\sum}\limits_{n\neq
m}\frac{p_nd_{n,m}}{(1+q_md_{n,m}\phi_n(\mathbf{q}))^2}}\quad \forall m
\end{equation}
where $\phi_m'(\mathbf{q})=\frac{-\phi_m(\mathbf{q})}{w_m+\frac{1}{N}\mathop{\sum}\limits_{n\neq
m}\frac{q_nd_{m,n}}{(1+q_nd_{m,n}\phi_m(\mathbf{q}))^2}} \;\forall m$.
\end{lemma}
\begin{proof}
The proof is given in Appendix B.
\end{proof}
We can see from (\ref{large_eq_9}) that the effective interference in the primal network equals the following:
$\frac{1}{N}\mathop{\sum}\limits_{n\neq m}\frac{p_nd_{n,m}}{(1+q_md_{n,m}\phi_n(\mathbf{q}))^2}$. In order
to establish the \textit{effective primal network}, we consider the following constructed power control problem conditioned on
$\hat{\mathbf{q}}$ and $\hat{\boldsymbol{\phi}}$:
\begin{equation}\label{large_eq_10}
\begin{array}{ll}
\mathrm{maximize} &
\mathop{\min}\limits_m\;\frac{\hat{p}_md_{m,m}\frac{\hat{\phi}_m^2}{-\hat{\phi}_m'}}{\beta_m\left(\sigma_m+\frac{1}{N}\mathop{\sum}\limits_{n\neq
m}\frac{\hat{p}_nd_{n,m}}{(1+\hat{q}_md_{n,m}\hat{\phi}_n)^2}\right)}\\
\mathrm{subject\;to}& \frac{1}{N}\mathbf{w}^{\mathsf{T}}\hat{\mathbf{p}}\leq \bar{P},\quad \hat{\mathbf{p}}>0\\
\mathrm{variables:} & \hat{\mathbf{p}}.
\end{array}
\end{equation}
Then, by defining the vector
$\mathbf{e}^{\mathsf{PN}}\triangleq\left(\frac{-\hat{\phi}_1'}{d_{1,1}\hat{\phi}_1^2},\cdots,\frac{-\hat{\phi}_{JK}'}{d_{JK,JK}\hat{\phi}_{JK}^2}\right)^{\mathsf{T}}$
and the nonnegative matrix $\mathbf{E}^{\mathsf{PN}}$ as
\begin{equation}\label{large_eq_11}
E_{m,n}^{\mathsf{PN}}=\left\{
\begin{array}{ll}
0,&\quad \mathrm{if}\;m=n\\ \frac{d_{n,m}}{(1+\hat{q}_md_{n,m}\hat{\phi}_n)^2},&\quad \mathrm{if}\;m\neq n\\
\end{array} \right.
\end{equation}
the objective function in (\ref{large_eq_10}) can be expressed compactly as
$\frac{\hat{p}_m}{\left(\mathrm{diag}(\boldsymbol{\beta}\circ\mathbf{e}^{\mathsf{PN}})\left(\frac{1}{N}\mathbf{E}^{\mathsf{PN}}\hat{\mathbf{p}}+\boldsymbol{\sigma}\right)\right)_m}$,
whose algebraic structure leads to the following eigenvalue problem in terms of the power $\hat{\mathbf{p}}^*$ and weighted asymptotic
$\mathsf{SINR}$ $\zeta^*$:
\begin{equation}\label{large_eq_12}
\frac{\hat{\mathbf{p}}^*}{\zeta^*}=\mathrm{diag}\left(\boldsymbol{\beta}\circ\mathbf{e}^{\mathsf{PN}}\right)\left(\mathbf{E}^{\mathsf{PN}}+(1/\bar{P})\boldsymbol{\sigma}\mathbf{w}^{\mathsf{T}}\right)\frac{\hat{\mathbf{p}}^*}{N}.
\end{equation}
By comparing with (\ref{finite_eq_4}), we can see that $\mathbf{E}^{\mathsf{PN}}$ can be regarded as the effective cross channel interference
matrix and the effective primal network can be characterized by the nonnegative matrix
$\mathrm{diag}\left(\boldsymbol{\beta}\circ\mathbf{e}^{\mathsf{PN}}\right)\left(\mathbf{E}^{\mathsf{PN}}+(1/\bar{P})\boldsymbol{\sigma}\mathbf{w}^{\mathsf{T}}\right)$.
Compared with the effective dual network, $\mathbf{E}^{\mathsf{PN}}$ is not explicitly dependent on $\hat{\mathbf{p}}$. In the following, we employ
Perron-Frobenius theory to propose a distributed algorithm to compute $\hat{\mathbf{p}}^*$ given $\hat{\mathbf{q}}$ and
$\hat{\boldsymbol{\phi}}$, which is given in Table \ref{tab_4} and is referred to as \textit{Algorithm D}.

\begin{table}[t]
\caption{Algorithm D: Computation of $\hat{\mathbf{p}}$ given $\hat{\mathbf{q}}$ and $\hat{\boldsymbol{\phi}}$}\label{tab_4} \centering
\rule{\linewidth}{0.1mm}
\begin{itemize}
\item Initialize arbitrary $\hat{\mathbf{p}}[0]\in\mathbb{R}_{++}^{JK\times 1}$ with given $\hat{\mathbf{q}}$ and $\hat{\boldsymbol{\phi}}$ such that $\frac{1}{N}\mathbf{w}^{\mathsf{T}}\hat{\mathbf{p}}[0]\leq\bar{P}$.
\end{itemize}
\begin{enumerate}
\item Update primal network power $\hat{\mathbf{p}}[\ell+1]$:
\begin{eqnarray*}
\hat{p}_m[\ell+1]\\ =&\frac{-\hat{\phi}_m'\beta_m}{\hat{\phi}_m^2d_{m,m}}\left(\sigma_m+\frac{1}{N}\mathop{\sum}\limits_{n\neq
m}\frac{\hat{p}_n[\ell]d_{n,m}}{(1+\hat{q}_md_{n,m}\hat{\phi}_n)^2}\right) \quad \forall m.
\end{eqnarray*}
\item Normalize $\hat{\mathbf{p}}[\ell+1]$:
\begin{equation*}
\hat{\mathbf{p}}[\ell+1]\leftarrow\frac{N\bar{P}}{\mathbf{w}^{\mathsf{T}}\hat{\mathbf{p}}[\ell+1]}\hat{\mathbf{p}}[\ell+1].
\end{equation*}
\end{enumerate}
\rule{\linewidth}{0.1mm}

\end{table}

\begin{theorem}\label{theorem_3}
For given $\hat{\mathbf{q}}$ and $\hat{\boldsymbol{\phi}}$, starting from any initial $\hat{\mathbf{p}}[0]$, the $\hat{\mathbf{p}}[\ell]$ in
$\textit{Algorithm D}$ converges geometrically fast to the optimal solution $\hat{\mathbf{p}}^*(\hat{\mathbf{q}},\hat{\boldsymbol{\phi}})$ of
(\ref{large_eq_10}).
\end{theorem}
\begin{proof}
The proof is given in Appendix A.
\end{proof}

\begin{table*}[t]
\caption{Algorithm E: Computation of $\hat{\mathbf{p}}$ and $\hat{\mathbf{q}}$ for multicell downlink}\label{tab_5} \centering
\rule{\linewidth}{0.1mm}
\begin{itemize}
\item Initialize arbitrary $\hat{\boldsymbol{\phi}}[0]\in\mathbb{R}_{++}^{JK\times 1}$, $\hat{\mathbf{p}}[0]\in\mathbb{R}_{++}^{JK\times 1}$, and $\hat{\mathbf{q}}[0]\in\mathbb{R}_{++}^{JK\times 1}$ such that $\frac{1}{N}\mathbf{w}^{\mathsf{T}}\hat{\mathbf{p}}[0]\leq\bar{P}$, and
$\frac{1}{N}\boldsymbol{\sigma}^{\mathsf{T}}\hat{\mathbf{q}}[0]\leq\bar{P}$.
\end{itemize}
\begin{enumerate}
\item Update dual network power $\hat{\mathbf{q}}[\ell+1]$:
\begin{equation*}
\hat{q}_m[\ell+1]=\frac{\beta_m}{d_{m,m}}\left(w_m+\frac{1}{N}\sum_{n\neq
m}\frac{\hat{q}_n[\ell]d_{m,n}}{1+\hat{q}_n[\ell]d_{m,n}\hat{\phi}_m[\ell]}\right) \quad \forall m.
\end{equation*}
\item Normalize $\hat{\mathbf{q}}[\ell+1]$:
\begin{equation*}
\hat{\mathbf{q}}[\ell+1]\leftarrow\frac{N\bar{P}}{\boldsymbol{\sigma}^{\mathsf{T}}\hat{\mathbf{q}}[\ell+1]}\hat{\mathbf{q}}[\ell+1].
\end{equation*}
\item Update $\hat{\boldsymbol{\phi}}[\ell+1]$:
\begin{equation*}
\hat{\phi}_m[\ell+1]=\frac{\beta_m}{d_{m,m}}\frac{1}{\hat{q}_m[\ell+1]} \quad \forall m.
\end{equation*}
\item Update $\hat{\boldsymbol{\phi}}'[\ell+1]$:
\begin{equation*}
\hat{\phi}_m'[\ell+1]=\frac{-\hat{\phi}_m[\ell+1]}{w_m+\frac{1}{N}\mathop{\sum}\limits_{n\neq
m}\frac{\hat{q}_n[\ell+1]d_{m,n}}{(1+\hat{q}_n[\ell+1]d_{m,n}\hat{\phi}_m[\ell+1])^2}} \quad \forall m.
\end{equation*}
\item Update primal network power $\hat{\mathbf{p}}[\ell+1]$:
\begin{equation*}
\hat{p}_m[\ell+1]=\frac{-\hat{\phi}_m'[\ell+1]\beta_m}{\hat{\phi}_m^2[\ell+1]d_{m,m}}\left(\sigma_m+\frac{1}{N}\mathop{\sum}\limits_{n\neq
m}\frac{\hat{p}_n[\ell]d_{n,m}}{(1+\hat{q}_m[\ell+1]d_{n,m}\hat{\phi}_n[\ell+1])^2}\right) \quad
\forall m.
\end{equation*}
\item Normalize $\hat{\mathbf{p}}[\ell+1]$:
\begin{equation*}
\hat{\mathbf{p}}[\ell+1]\leftarrow\frac{N\bar{P}}{\mathbf{w}^{\mathsf{T}}\hat{\mathbf{p}}[\ell+1]}\hat{\mathbf{p}}[\ell+1].
\end{equation*}
\end{enumerate}
\rule{\linewidth}{0.1mm}

\end{table*}

Now, by combining \textit{Algorithm B, C, and D} that have respectively treated $\hat{\boldsymbol{\phi}}$, $\hat{\mathbf{q}}$ and
$\hat{\mathbf{p}}$ separately, a single timescale algorithm is given in Table \ref{tab_5} and is referred to as \textit{Algorithm E}. Even
though this algorithm that computes the asymptotic power is iterative, it only requires statistical channel information and thus the asymptotic
power is updated at a slower timescale. Then for each instantaneous time, the asymptotic primal network power $\hat{\mathbf{p}}^*$ is used for
the downlink transmission, and the asymptotic dual network power $\hat{\mathbf{q}}^*$ is employed to non-iteratively obtain the instantaneous
beamforming matrix $\hat{\mathbb{U}}^*$ with local CSI as $\hat{\mathbf{u}}_m^*(\hat{\mathbf{q}}^*)=\frac{(\sum_{n\neq
m}\frac{\hat{q}_n^*}{N}\mathbf{h}_{m,n}\mathbf{h}_{m,n}^{\dag}+w_m\mathbf{I})^{-1}\mathbf{h}_{m,m}}{\|(\sum_{n\neq
m}\frac{\hat{q}_n^*}{N}\mathbf{h}_{m,n}\mathbf{h}_{m,n}^{\dag}+w_m\mathbf{I})^{-1}\mathbf{h}_{m,m}\|}$. In this way, by leveraging the
asymptotic property in the large scale system, no instantaneous power update is required in the coordinated cluster to jointly optimize power
control and beamformer. To draw connection with the finite system analysis, we summarize the results obtained by the nonlinear Perron-Frobenius
theory in Table \ref{tab_6}.

\renewcommand{\arraystretch}{1.5}

\begin{table*}[t]\scriptsize
\caption{Nonlinear Perron-Frobenius characterization: finite system of the dual network (second row); finite system of the primal network (third
row); large system of the dual network (fourth row); large system of the primal network (fifth row).}\label{tab_6} \centering
\begin{tabular}{|c|c|c|}
\hline \textit{Concave Self-Mapping} & \textit{Perron Eigenvalue} & \textit{Perron Eigenvector}\\
\hline\hline
$\mathrm{diag}(\boldsymbol{\beta}\circ\mathbf{g}(\mathbb{U}))\left(\frac{1}{N}\mathbf{F}^{\mathsf{T}}(\mathbb{U})\mathbf{q}(\mathbb{U})+\mathbf{w}\right),\;\forall
\mathbb{U}$ &
$\rho\left(\mathrm{diag}(\boldsymbol{\beta}\circ\mathbf{g}(\mathbb{U}))\left(\mathbf{F}^{\mathsf{T}}(\mathbb{U})+(1/\bar{P})\mathbf{w}\boldsymbol{\sigma}^{\mathsf{T}}\right)\right)$
&
$\mathbf{x}\left(\mathrm{diag}(\boldsymbol{\beta}\circ\mathbf{g}(\mathbb{U}))\left(\mathbf{F}^{\mathsf{T}}(\mathbb{U})+(1/\bar{P})\mathbf{w}\boldsymbol{\sigma}^{\mathsf{T}}\right)\right)$\\
\hline
$\mathrm{diag}(\boldsymbol{\beta}\circ\mathbf{g}(\mathbb{U}))\left(\frac{1}{N}\mathbf{F}(\mathbb{U})\mathbf{p}(\mathbb{U})+\boldsymbol{\sigma}\right),\;\forall
\mathbb{U}$ &
$\rho\left(\mathrm{diag}(\boldsymbol{\beta}\circ\mathbf{g}(\mathbb{U}))\left(\mathbf{F}(\mathbb{U})+(1/\bar{P})\boldsymbol{\sigma}\mathbf{w}^{\mathsf{T}}\right)\right)$
&
$\mathbf{x}\left(\mathrm{diag}(\boldsymbol{\beta}\circ\mathbf{g}(\mathbb{U}))\left(\mathbf{F}(\mathbb{U})+(1/\bar{P})\boldsymbol{\sigma}\mathbf{w}^{\mathsf{T}}\right)\right)$\\
\hline\hline
$\mathrm{diag}(\boldsymbol{\beta}\circ\mathbf{e}^{\mathsf{DN}})\left(\frac{1}{N}\mathbf{E}^{\mathsf{DN}}(\hat{\mathbf{q}})\hat{\mathbf{q}}+\mathbf{w}\right)$
&
$\rho\left(\mathrm{diag}(\boldsymbol{\beta}\circ\mathbf{e}^{\mathsf{DN}})\left(\mathbf{E}^{\mathsf{DN}}(\hat{\mathbf{q}})+(1/\bar{P})\mathbf{w}\boldsymbol{\sigma}^{\mathsf{T}}\right)\right)$
&
$\mathbf{x}\left(\mathrm{diag}(\boldsymbol{\beta}\circ\mathbf{e}^{\mathsf{DN}})\left(\mathbf{E}^{\mathsf{DN}}(\hat{\mathbf{q}})+(1/\bar{P})\mathbf{w}\boldsymbol{\sigma}^{\mathsf{T}}\right)\right)$\\
\hline
$\mathrm{diag}(\boldsymbol{\beta}\circ\mathbf{e}^{\mathsf{PN}})\left(\frac{1}{N}\mathbf{E}^{\mathsf{PN}}\hat{\mathbf{p}}+\boldsymbol{\sigma}\right)$
&
$\rho\left(\mathrm{diag}(\boldsymbol{\beta}\circ\mathbf{e}^{\mathsf{PN}})\left(\mathbf{E}^{\mathsf{PN}}+(1/\bar{P})\boldsymbol{\sigma}\mathbf{w}^{\mathsf{T}}\right)\right)$
&
$\mathbf{x}\left(\mathrm{diag}(\boldsymbol{\beta}\circ\mathbf{e}^{\mathsf{PN}})\left(\mathbf{E}^{\mathsf{PN}}+(1/\bar{P})\boldsymbol{\sigma}\mathbf{w}^{\mathsf{T}}\right)\right)$\\
\hline
\end{tabular}
\end{table*}

\textit{Discussion of Complexity:} It is important to note that even though \textit{Algorithm A} and \textit{Algorithm E} are both discrete time
algorithms, their operating timescales as well as the implementation complexities are vastly different (we use indices $\kappa$ and $\ell$ to
differentiate them). In \textit{Algorithm A}, the power update is on the order of milliseconds to track the instantaneous channel effect. Thus,
this algorithm requires a large amount of instantaneous power update to compute the optimal solution. In contrast, the power update in
\textit{Algorithm E} relies only on statistical channel information. Therefore, this algorithm operates on the order of tens of seconds or more
(at the same timescale as the variation of the long-term channel statistics) and thus the implementation complexity is greatly reduced.

\section{Numerical Results}\label{numerical}
In this section, we conduct a numerical study to support the analysis. We consider a three-cell cluster, i.e., $J=3$, wherein the three base
stations jointly perform power control and coordinated beamforming. The path loss (in dB) model in \cite{3gpp10} is assumed with
$15.3+37.6\log_{10} d$ for distance $d$ in meters and a log-normal shadowing with standard deviation of $8$ dB is employed. The noise power
spectral density is set to $-162$ dBm/Hz. The radius of each base station is set to be $1.5$ km, and a $15$ dBi antenna gain is assumed. For
illustration purpose, the coordinated cluster is subject to a total power constraint, i.e., $\mathbf{w}=\mathbf{1}$, and each user possesses the
same priority of service, i.e., $\boldsymbol{\beta}=\mathbf{1}$. The total power constraint $\bar{P}$ is assumed to equal $10$ Watt.

\begin{figure}[t]
\centering
    \includegraphics[width=1.0\linewidth]{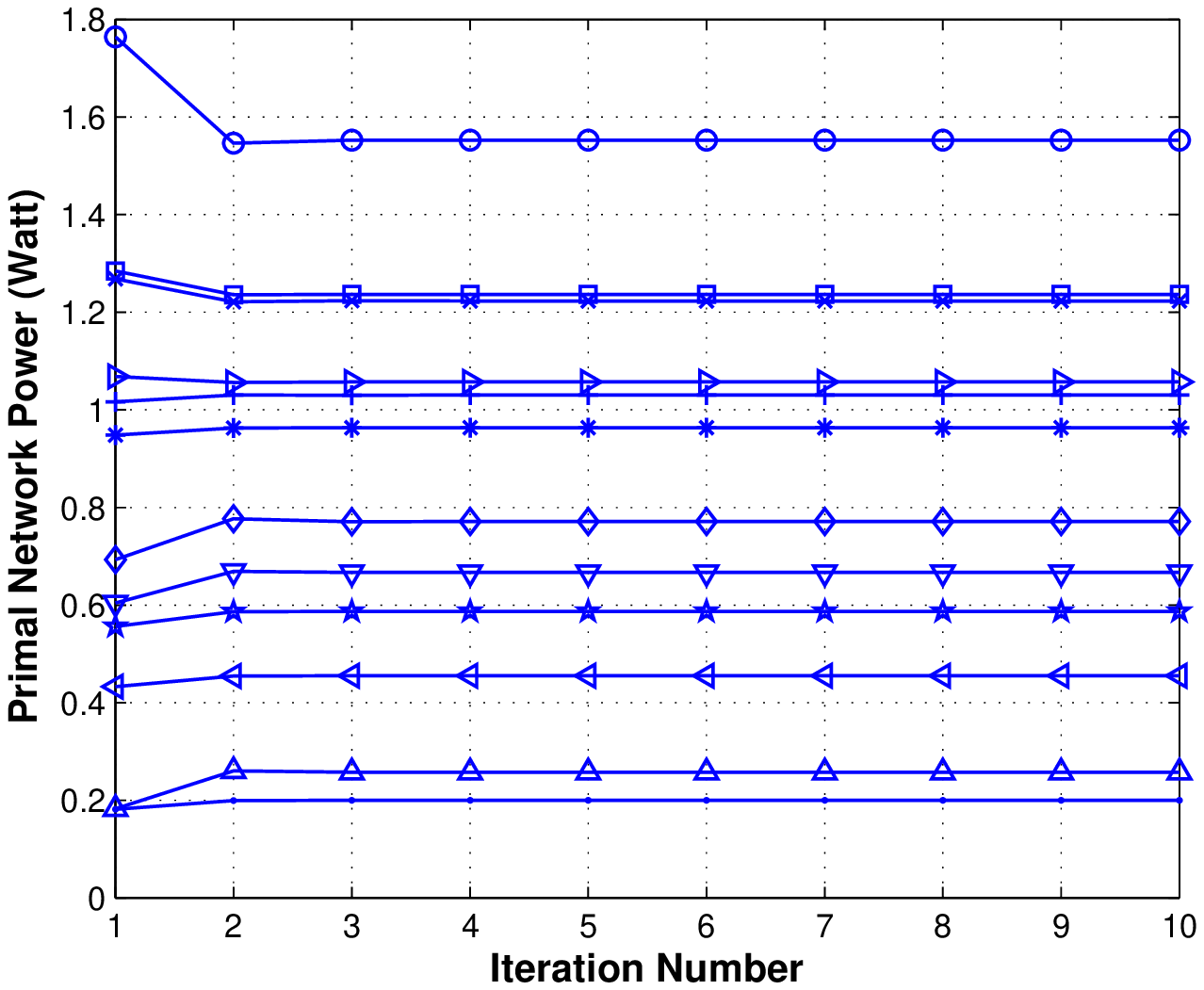}
\caption{Convergence plot of the primal network power in a finite system setting employing Algorithm A: ($N=4$, $K=4$, $J=3$, $\bar{P}=10$
Watt). Different marker types represent different users.} \label{fig_1}
\end{figure}

\begin{figure}[t]
\centering
    \includegraphics[width=1.0\linewidth]{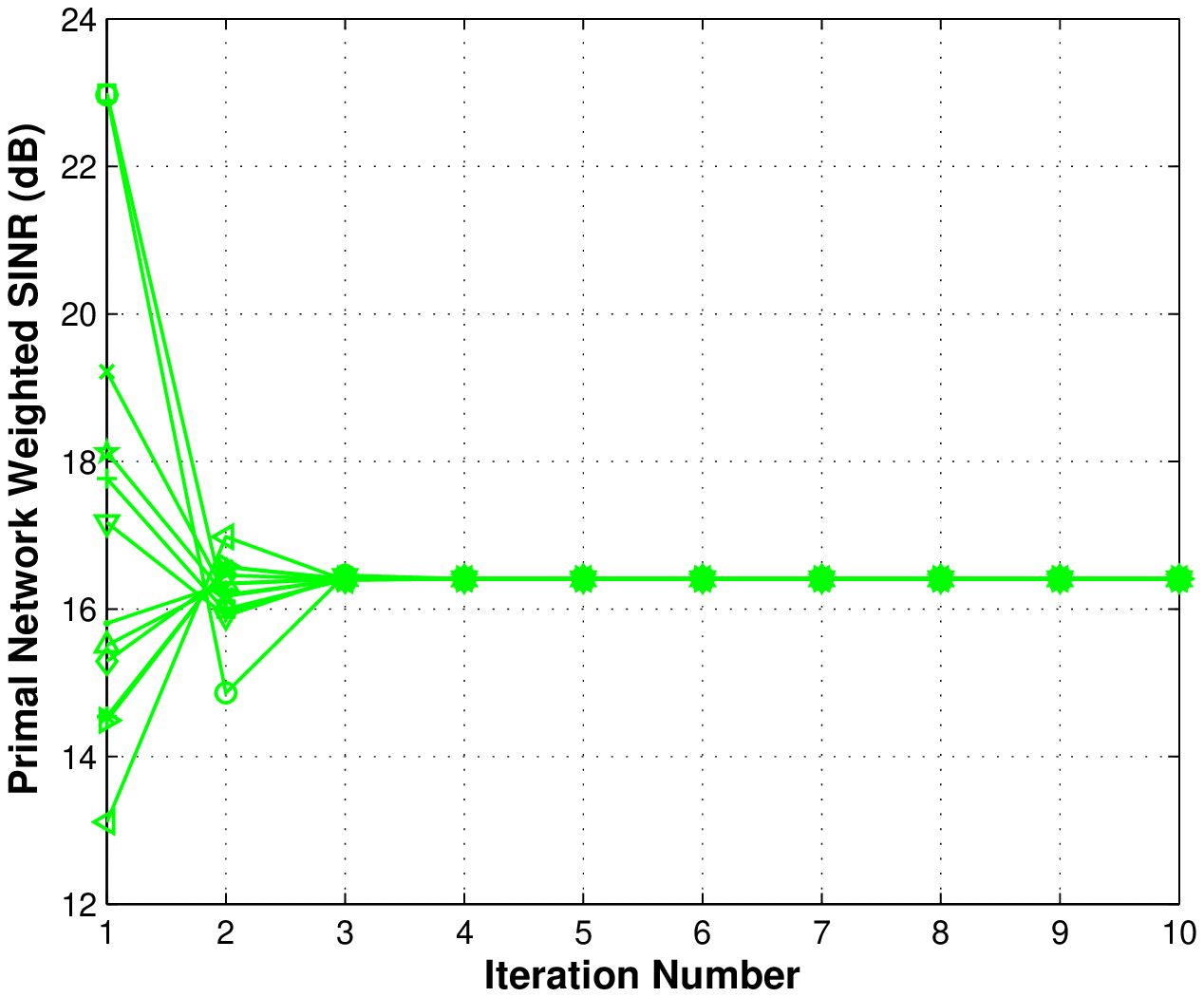}
\caption{Convergence plot of the primal network weighted $\mathsf{SINR}$ in a finite system setting employing Algorithm A: ($N=4$, $K=4$, $J=3$
$\bar{P}=10$ Watt). Different marker types represent different users.} \label{fig_2}
\end{figure}

Firstly, a finite system setting is considered. Each base station is assumed to be equipped with $N=4$ antennas and serves $K=4$ randomly
located users simultaneously. For one channel realization, the coordinated cluster utilizes \textit{Algorithm A} to iteratively obtain the
optimal beamformer and optimal power. Fig. \ref{fig_1} demonstrates the convergence plot of the primal network power for different users. It is
observed that for this channel realization, the power converges to its optimal value within $3$ runs of iteration. This demonstrates the
geometrically fast convergent property of \textit{Algorithm A}. Extensive numerical evaluations show that it converges typically within $10$
runs of iteration for different channel realizations. In Fig. \ref{fig_2}, the convergence plot of the primal network weighted $\mathsf{SINR}$
for different users is shown. Since the system metric of the coordinated cluster is to maintain fairness across users, each user's optimal
$\mathsf{SINR}$ ($\mathbf{w}=\mathbf{1}$) would converge to the same value for a given channel realization, which is verified in Fig.
\ref{fig_2}.

\begin{figure}[t]
\centering
    \includegraphics[width=1.0\linewidth]{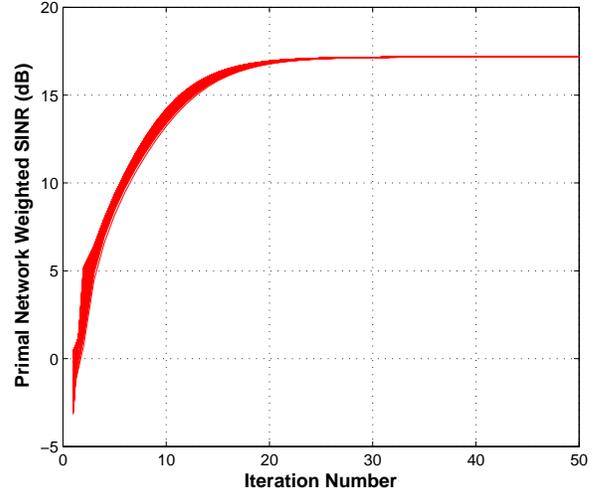}
\caption{Convergence plot of the primal network weighted $\mathsf{SINR}$ in a large system setting employing Algorithm E: ($N=50$, $K=40$,
$J=3$, $\bar{P}=10$ Watt). The $\mathsf{SINR}$ for different users are not differentiated, but use the same line style for illustration.}
\label{fig_3}
\end{figure}

\begin{figure}[t]
\centering
    \includegraphics[width=1.0\linewidth]{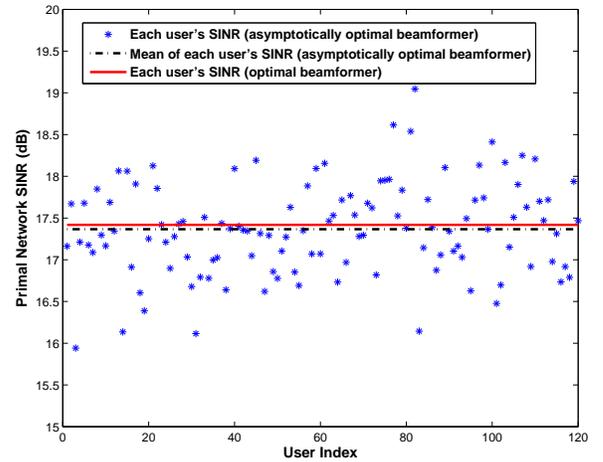}
\caption{The achieved primal network $\mathsf{SINR}$ for each individual user using the asymptotically optimal beamformer in a large system
setting for one channel realization: ($N=50$, $K=40$, $J=3$, $\bar{P}=10$ Watt). The mean of the achieved $\mathsf{SINR}$ using asymptotically
optimal beamformer averaged over users and the the achieved $\mathsf{SINR}$ using the optimal beamformer are illustrated for comparison.}
\label{fig_4}
\end{figure}

Next, a large system setting is considered with $N=50$ and $K=40$. For a given geometry, the asymptotic $\mathsf{SINR}$ of the primal network is
of interest, whose convergence plot is shown in Fig. \ref{fig_3} by employing \textit{Algorithm E}. The $\mathsf{SINR}$'s of each user are not
differentiated, and uses the same line of type for illustration. Note that the converged value does not depend on the channel realization.
However, it depends on the user geometry, namely the large scale channel effects, which means different user geometries would lead to different
deterministic equivalents for the optimal $\mathsf{SINR}$ in the large system. Fig. \ref{fig_4} considers the use of asymptotic result. The
asymptotic primal network power is utilized for downlink transmission, and the asymptotic dual network power is leveraged to non-iteratively
determine the instantaneous beamformer. The achieved $\mathsf{SINR}$'s for different users using the determined beamformer are shown, along with
their mean and the achieved $\mathsf{SINR}$ using the optimal beamformer obtained via \textit{Algorithm A}, for one channel realization. It is
observed that the $\mathsf{SINR}$'s of different users employing the asymptotically optimal beamformer fluctuate around the optimal one, with
the mean close to the optimal $\mathsf{SINR}$. Therefore, by using \textit{Algorithm E} to obtain the asymptotically optimal beamformer, the
max-min fairness across users can be achieved in the asymptotic sense.

\begin{figure}[t]
\centering
    \includegraphics[width=1.0\linewidth]{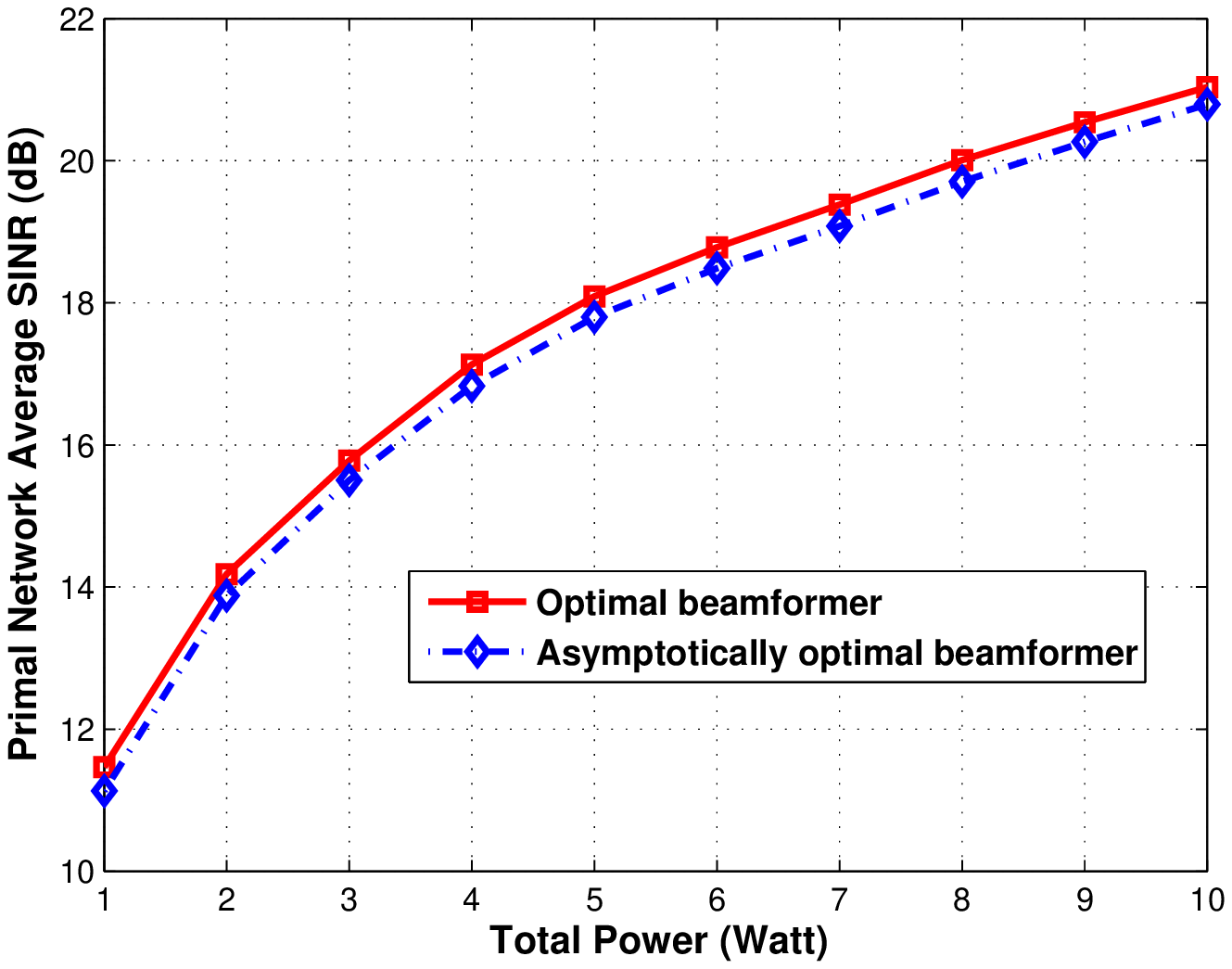}
\caption{Comparison of the average achieved primal network $\mathsf{SINR}$ using asymptotically optimal beamformer and the optimal beamformer in
a finite system setting with respect to different values of the power constraint $\bar{P}$. The averaging is performed over different geometries
of users and different channel realizations: ($N=4$, $K=3$, $J=3$).} \label{fig_5}
\end{figure}

Finally, in Fig. \ref{fig_5}, we consider the use of the asymptotic result in a finite system with $N=4$ and $K=3$ and demonstrate the
comparison of the average $\mathsf{SINR}$ using optimal beamformer and the asymptotically optimal beamformer with respect to the variation of
the total power constraint $\bar{P}$. Herein, the averaging is over the user geometries, and for a given user geometry, different channel
realizations are drawn. It can be seen from Fig. \ref{fig_5} that the performance of applying asymptotic result holds well for finite system
setting. Accordingly, in a practical system with limited backhaul constraint, the asymptotically optimal power and beamformer can be developed
and leveraged to reduce the implementation complexity and approach the optimal performance in the asymptotic sense.

\section{Conclusion}\label{conclusion}
In this paper, we consider a joint optimization of beamforming and power control in a coordinated multicell downlink and employ the max-min
formulation to enforce egalitarian fairness across users. The network duality is interpreted via a nonlinear Perron-Frobenius theoretic
characterization and utilized to design a distributed algorithm to obtain the optimal solution. The iterative algorithm requires instantaneous
power update through the limited backhaul and does not scale well in a large system setting. In order to design an algorithm that only utilizes
channel statistics, we leverage random matrix theory to derive deterministic equivalents for the optimal $\mathsf{SINR}$ expression, and utilize
the nonlinear Perron-Forbenius theory to establish the notion of effective network and propose a fast convergent algorithm. The asymptotically
optimal solution enables a non-iterative approach to compute the instantaneous beamformer and thus requires no instantaneous information
exchange across the coordinated cluster. This paper assumes an independent channel model and utilizes a weighted sum power constraint.
Investigating the impact of practical issues on algorithm design in a large system setting, such as channel estimation error and per cell power
constraint are interesting directions of future work.


%

\appendices
\section{}\label{appenA}

\textit{Proof of Theorem \ref{theorem_1}:} The key step to the proof is to establish the convergence property of the dual network power
$\mathbf{q}$ via a nonlinear Perron-Frobenius theory in \cite{cai11}. The relationship between $\mathbf{q}^*$ and the optimal weighted
$\mathsf{SINR}$ $\tau^*$ is of interest, and can be obtained by substituting the optimal MVDR beamformer:
\begin{equation}\label{appen_eq_1}
\frac{q_m^*}{N\tau^*}=\frac{\beta_m}{\mathbf{h}_{m,m}^{\dag}(\sum_{n\neq
m}\frac{q_n^*}{N}\mathbf{h}_{m,n}\mathbf{h}_{m,n}^{\dag}+w_m\mathbf{I})^{-1}\mathbf{h}_{m,m}} \quad \forall m.
\end{equation}
Thus the mapping $\mathcal{I}^{(1)}(\cdot): \mathbb{R}_+^{JK\times 1}\rightarrow\mathbb{R}_+^{JK\times 1}$ can be defined by the following
equation: $\mathcal{I}_m^{(1)}(\mathbf{q}^*)\triangleq\frac{\beta_m}{\mathbf{h}_{m,m}^{\dag}(\sum_{n\neq
m}\frac{q_n^*}{N}\mathbf{h}_{m,n}\mathbf{h}_{m,n}^{\dag}+w_m\mathbf{I})^{-1}\mathbf{h}_{m,m}}$. It can be shown using the same technique in
\cite{cai11} that $\mathcal{I}^{(1)}(\cdot)$ is a concave self-mapping of $\mathbf{q}^*$. Also, for the dual network, the weighted sum power
constraint $\frac{1}{N}\boldsymbol{\sigma}^{\mathsf{T}}\mathbf{q}^*=\bar{P}$ induces a norm on $\mathbb{R}_+^{JK\times 1}$ defined by
$\|\mathbf{q}^*\|_{\mathsf{DN}}\triangleq (N/\bar{P})\sum_m \sigma_m q_m^*$. By applying \cite[Theorem 1]{krause01}, starting from any initial
point $\mathbf{q}[0]$, the fixed-point iteration (step $1$ and step $2$ of \textit{Algorithm A}) converges geometrically fast to the optimal
solution $\mathbf{q}^*$ for the eigenvalue problem (\ref{appen_eq_1}). The optimal beamforming matrix $\mathbb{U}^*$ is unique and can be
computed by substituting the optimal dual network power $\mathbf{q}^*$ into the MVDR beamformer (\ref{finite_eq_8}) for each user (step $3$ of
\textit{Algorithm A}). For the primal network power $\mathbf{p}$, the induced norm on $\mathbb{R}_+^{JK\times 1}$ is established by the weighted
power constraint $\frac{1}{N}\mathbf{w}^{\mathsf{T}}\mathbf{p}^*=\bar{P}$ as: $\|\mathbf{p}^*\|_{\mathsf{PN}}\triangleq (N/\bar{P})\sum_m w_m
p_m^*$. Therefore, by using the same line of argument for the dual network with the optimal beamforming matrix $\mathbb{U}^*$, the fixed-point
iteration (step $4$ and step $5$ of \textit{Algorithm A}) converges geometrically fast to the optimal solution $\mathbf{p}^*$ for the eigenvalue
problem (\ref{finite_eq_4}) with any initial point $\mathbf{p}[0]$. This completes the proof of Theorem \ref{theorem_1}.

\medskip

\textit{Proof of Theorem \ref{theorem_2}:} For a given $\hat{\boldsymbol{\phi}}$, the nonlinear eigenvalue problem in (\ref{large_eq_7}) enables
us to define the mapping $\mathcal{I}^{(3)}(\cdot): \mathbb{R}_+^{JK\times 1}\rightarrow\mathbb{R}_+^{JK\times 1}$ as:
$\mathcal{I}_m^{(3)}(\hat{\mathbf{q}})\triangleq\frac{\beta_m}{d_{m,m}}\left(w_m+\frac{1}{N}\sum_{n\neq
m}\frac{\hat{q}_nd_{m,n}}{1+\hat{q}_nd_{m,n}\hat{\phi}_m}\right)$. Since the function $\frac{x}{1+x}$ is strictly concave in
$x\in\mathbb{R}_{++}$, the mapping $\mathcal{I}_m^{(3)}(\hat{\mathbf{q}})$ is a summation of strictly concave functions in $\hat{\mathbf{q}}$
and thus is a concave self-mapping in $\hat{\mathbf{q}}$. Then using the norm $\|\mathbf{q}\|_{\mathsf{DN}}$ in Appendix A and applying
\cite[Theorem 1]{krause01}, the fixed-point iteration (step $1$ and $2$ of \textit{Algorithm C}) converges geometrically fast to
$\hat{\mathbf{q}}^*(\hat{\boldsymbol{\phi}})$ for the eigenvalue problem (\ref{large_eq_7}).

\medskip

\textit{Proof of Theorem \ref{theorem_3}:} For given $\hat{\boldsymbol{\phi}}$ and $\hat{\mathbf{q}}$, the eigenvalue problem in
(\ref{large_eq_12}) enables us to define the mapping $\mathcal{I}^{(4)}(\cdot): \mathbb{R}_+^{JK\times 1}\rightarrow\mathbb{R}_+^{JK\times 1}$
as:
$\mathcal{I}_m^{(4)}(\hat{\mathbf{p}})\triangleq\frac{-\hat{\phi}_m'\beta_m}{\hat{\phi}_m^2d_{m,m}}\left(\sigma_m+\frac{1}{N}\mathop{\sum}\limits_{n\neq
m}\frac{\hat{p}_nd_{n,m}}{(1+\hat{q}_md_{n,m}\hat{\phi}_n)^2}\right)$. It can be easily seen that the mapping
$\mathcal{I}_m^{(4)}(\hat{\mathbf{p}})$ is affine, thus it is a concave self-mapping in $\hat{\mathbf{p}}$. Then using the norm
$\|\mathbf{p}\|_{\mathsf{PN}}$ in Appendix A and applying \cite[Theorem 1]{krause01}, the fixed-point iteration (step $1$ and $2$ of
\textit{Algorithm D}) converges geometrically fast to $\hat{\mathbf{p}}^*(\hat{\boldsymbol{\phi}},\hat{\mathbf{q}})$ for the eigenvalue problem
(\ref{large_eq_12}).

\section{}\label{appenB}
\textit{Useful Results from Random Matrix Theory:} We reproduce the following theorem \cite{kammoun09, lakshminaryana10, zakhour11} that will be
employed to prove Lemma \ref{lemma_1} and Lemma \ref{lemma_3}.

\begin{theorem}\label{theorem_4}
(Theorem $2$ in \cite{kammoun09}) Consider an $\tilde{N}\times\tilde{n}$ random matrix $\mathbf{Y}=\left(Y_{i,j}\right)_{i=1,j=1}^{\tilde{N},
\tilde{n}}$ where the entries are given by: $Y_{i,j}=\frac{\tilde{\sigma}_{i,j}}{\sqrt{\tilde{n}}}X_{i,j}$, the $X_{i,j}$ being independent and
identically distributed (i.i.d.), with the following assumptions hold:

\textit{A1}: The complex random variables $X_{i,j}$ are i.i.d. with $\mathbb{E}[X_{i,j}]=0$, $\mathbb{E}[X_{i,j}^2]=0$,
$\mathbb{E}[|X_{i,j}|^2]=1$, and $\mathbb{E}[|X_{i,j}|^8]<\infty$.

\textit{A2}: There exists a real number $\tilde{\sigma}_{\mathrm{max}}<\infty$ such that: $\mathop{\sup}\limits_{\tilde{n}\geq
1}\mathop{\max}\limits_{\substack{1\leq i\leq \tilde{N}\\1\leq j\leq \tilde{n}}}|\tilde{\sigma}_{i,j}|\leq \tilde{\sigma}_{\mathrm{max}}$.

There exists a deterministic $\tilde{N}\times\tilde{N}$ matrix-valued function
$\mathbf{\Psi}(z)=\mathrm{diag}(\psi_1(z),\ldots,\psi_{\tilde{N}}(z))$ analytic in $\mathbb{C}-\mathbb{R}_+$ such that:
\begin{equation}\label{appen_eq_6}
\frac{1}{\tilde{N}}\mathsf{Tr}\left(\mathbf{Y}\mathbf{Y}^{\dag}-z\mathbf{I}_{\tilde{N}}\right)^{-1}-\frac{1}{\tilde{N}}\mathsf{Tr}(\mathbf{\Psi}(z))\mathop{\longrightarrow}\limits^{a.s.}0
\quad \mathrm{for} \; z\in \mathbb{C}-\mathbb{R}_+
\end{equation}
whose elements are the unique solutions of the deterministic system of $\tilde{N}+\tilde{n}$ equations:
\begin{align}
\psi_i(z)&=\frac{-1}{z\left(1+\frac{1}{\tilde{n}}\sum_{j=1}^{\tilde{n}}\tilde{\sigma}_{i,j}^2\tilde{\psi}_j(z)\right)} \quad \mathrm{for}\; 1\leq i\leq \tilde{N}\notag\\
\label{appen_eq_7} \tilde{\psi}_j(z)&=\frac{-1}{z\left(1+\frac{1}{\tilde{n}}\sum_{i=1}^{\tilde{N}}\tilde{\sigma}_{i,j}^2\psi_i(z)\right)} \quad
\mathrm{for}\; 1\leq j\leq \tilde{n}
\end{align}
such that $\frac{1}{\tilde{N}}\mathsf{Tr}(\mathbf{\Psi}(z))$ is the Stieltjes transform \cite{tulino04} of a probability measure.
\end{theorem}

\medskip

\begin{figure*}
\begin{equation}\label{appen_eq_2}
\gamma_m^{\mathsf{DN}}(\mathbf{q})-\frac{q_md_{m,m}}{N}\mathsf{Tr}\left(\left(\sum_{n\neq
m}\frac{q_nd_{m,n}}{N}\tilde{\mathbf{h}}_{m,n}\tilde{\mathbf{h}}_{m,n}^{\dag}+w_m\mathbf{I}\right)^{-1}\right)\mathop{\longrightarrow}\limits^{a.s.}0.
\end{equation}
\end{figure*}

\begin{figure*}
\begin{equation}\label{appen_eq_3}
\frac{1}{N}|\tilde{\mathbf{h}}_{m,m}^{\dag}\mathbf{u}_m^*|^2=\frac{\left(\frac{1}{N}\tilde{\mathbf{h}}_{m,m}^{\dag}\left(\sum_{n\neq
m}\frac{q_nd_{m,n}}{N}\tilde{\mathbf{h}}_{m,n}\tilde{\mathbf{h}}_{m,n}^{\dag}+w_m\mathbf{I}\right)^{-1}\tilde{\mathbf{h}}_{m,m}\right)^{2}}{\frac{1}{N}\tilde{\mathbf{h}}_{m,m}^{\dag}\left(\sum_{n\neq
m}\frac{q_nd_{m,n}}{N}\tilde{\mathbf{h}}_{m,n}\tilde{\mathbf{h}}_{m,n}^{\dag}+w_m\mathbf{I}\right)^{-2}\tilde{\mathbf{h}}_{m,m}}.
\end{equation}
\end{figure*}

\begin{figure*}[th]
\begin{equation}\label{appen_eq_4}
\frac{1}{N}|\tilde{\mathbf{h}}_{n,m}^{\dag}\mathbf{u}_n^*|^2=\frac{\frac{1}{N}\tilde{\mathbf{h}}_{n,m}^{\dag}\left(\mathop{\sum}\limits_{\jmath\neq
n}\frac{q_{\jmath}d_{n,\jmath}}{N}\tilde{\mathbf{h}}_{n,\jmath}\tilde{\mathbf{h}}_{n,\jmath}^{\dag}+w_n\mathbf{I}\right)^{-1}\tilde{\mathbf{h}}_{n,n}\tilde{\mathbf{h}}_{n,n}^{\dag}\left(\mathop{\sum}\limits_{\jmath\neq
n}\frac{q_{\jmath}d_{n,\jmath}}{N}\tilde{\mathbf{h}}_{n,\jmath}\tilde{\mathbf{h}}_{n,\jmath}^{\dag}+w_n\mathbf{I}\right)^{-1}\tilde{\mathbf{h}}_{n,m}}{\frac{1}{N}\tilde{\mathbf{h}}_{n,n}^{\dag}\left(\mathop{\sum}\limits_{\jmath\neq
n}\frac{q_{\jmath}d_{n,\jmath}}{N}\tilde{\mathbf{h}}_{n,\jmath}\tilde{\mathbf{h}}_{n,\jmath}^{\dag}+w_n\mathbf{I}\right)^{-2}\tilde{\mathbf{h}}_{n,n}}.
\end{equation}
\end{figure*}

\begin{figure*}
\begin{equation}\label{appen_eq_5}
\frac{\frac{1}{N}\tilde{\mathbf{h}}_{n,m}^{\dag}\left(\mathop{\sum}\limits_{\jmath\neq
m,n}\frac{q_{\jmath}d_{n,\jmath}}{N}\tilde{\mathbf{h}}_{n,\jmath}\tilde{\mathbf{h}}_{n,\jmath}^{\dag}+w_n\mathbf{I}\right)^{-1}\tilde{\mathbf{h}}_{n,n}\tilde{\mathbf{h}}_{n,n}^{\dag}\left(\mathop{\sum}\limits_{\jmath\neq
m,n}\frac{q_{\jmath}d_{n,\jmath}}{N}\tilde{\mathbf{h}}_{n,\jmath}\tilde{\mathbf{h}}_{n,\jmath}^{\dag}+w_n\mathbf{I}\right)^{-1}\tilde{\mathbf{h}}_{n,m}}{\left(1+\frac{q_md_{n,m}}{N}\tilde{\mathbf{h}}_{n,m}^{\dag}\left(\mathop{\sum}\limits_{\jmath\neq
m,n}\frac{q_{\jmath}d_{n,\jmath}}{N}\tilde{\mathbf{h}}_{n,\jmath}\tilde{\mathbf{h}}_{n,\jmath}^{\dag}+w_n\mathbf{I}\right)^{-1}\tilde{\mathbf{h}}_{n,m}\right)^2}.
\end{equation}
\rule{\linewidth}{0.1pt}
\end{figure*}

\textit{Proof of Lemma \ref{lemma_1}:} The technique to establish the deterministic equivalent for $\gamma_m^{\mathsf{DN}}(\mathbf{q})$ lies in
the asymptotic behavior of the empirical distribution of the eigenvalue for $\left(\sum_{n\neq
m}\frac{q_nd_{m,n}}{N}\tilde{\mathbf{h}}_{m,n}\tilde{\mathbf{h}}_{m,n}^{\dag}+w_m\mathbf{I}\right)^{-1}$. This uplink problem for the equal
power system has been addressed in \cite{tse99}, and the general treatment using the notion of variance profiles for random matrices is provided
in \cite{kammoun09}. Applying \cite[Lemma 2.7]{bai98} yields (27).

Since the separable variance profile for the Gram matrix $\sum_{n\neq
m}\frac{q_nd_{m,n}}{N}\tilde{\mathbf{h}}_{m,n}\tilde{\mathbf{h}}_{m,n}^{\dag}$ is characterized by the optimal power $q_n$ and the large-scale
channel effects $d_{m,n}$, there exist a deterministic equivalent for the Stieltjes transform \cite{tulino04} of this Gram matrix. In order to
invoke Theorem \ref{theorem_4}, the channel model needs to satisfy the two assumptions (i.e., $\textit{A1}$ and $\textit{A2}$) described above.
Note that the channel model in (\ref{large_eq_1}) constitutes a special case of the channel model assumed in \cite{hachem07, kammoun09},
therefore the matrices considered satisfy the two necessary assumptions. Employing Theorem \ref{theorem_4} generates the fixed-point equation
for $\gamma_m^{\mathsf{DN}}(\mathbf{q})$ in (\ref{large_eq_3}).

\medskip

\textit{Proof of Lemma \ref{lemma_2}:} For a given $\hat{\mathbf{q}}$, define the following mapping:
$\mathcal{I}_m^{(2)}({\hat{\phi_m}})\triangleq\frac{1}{w_m+\frac{1}{N}\mathop{\sum}\limits_{n\neq
m}\frac{\hat{q}_nd_{m,n}}{1+\hat{q}_nd_{m,n}\hat{\phi}_m}}$. The idea for proving this lemma is to use the standard interference function
framework \cite{yates95}. It is straightforward to check that the positivity and monotonicity conditions in \cite{yates95} hold for
$\mathcal{I}_m^{(2)}({\hat{\phi_m}})$. Also, for all $\varepsilon>1$, we have
$\frac{1}{\frac{w_m}{\varepsilon}+\frac{1}{N}\mathop{\sum}\limits_{n\neq
m}\frac{\hat{q}_nd_{m,n}}{\varepsilon+\hat{q}_nd_{m,n}\varepsilon\hat{\phi}_m}}>\frac{1}{w_m+\frac{1}{N}\mathop{\sum}\limits_{n\neq
m}\frac{\hat{q}_nd_{m,n}}{1+\hat{q}_nd_{m,n}\varepsilon\hat{\phi}_m}}$, which establishes the scalability condition in \cite{yates95}. Since the
mapping is a standard interference function, the convergence result follows from \cite{yates95}, thus completing the proof of Lemma
\ref{lemma_2}.

\medskip

\textit{Proof of Lemma \ref{lemma_3}:} The expression for $\Gamma_m^{\mathsf{PN}}(\mathbf{p})$ is given in (\ref{large_eq_8}), and the optimal
beamformer $\mathbf{u}_m^*$ is the MVDR beamformer in (\ref{finite_eq_8}). The asymptotic approximations for
$\frac{1}{N}|\tilde{\mathbf{h}}_{m,m}^{\dag}\mathbf{u}_m^*|^2$ and $\frac{1}{N}|\tilde{\mathbf{h}}_{n,m}^{\dag}\mathbf{u}_n^*|^2$ need to be
determined. The expression for $\frac{1}{N}|\tilde{\mathbf{h}}_{m,m}^{\dag}\mathbf{u}_m^*|^2$ can be further expanded as (28).

Employing Theorem \ref{theorem_4}, the numerator of (\ref{appen_eq_3}) converges almost surely to $\phi_m^2(\mathbf{q})$. In order to obtain the
deterministic equivalent for the denominator, the dependence of $\phi_m(\mathbf{q})$ on the noise variance $w_m$ can be made explicit, i.e.,
$\phi_m(\mathbf{q})=\phi_m(\mathbf{q},x)|_{x=w_m}$. Then, by employing the differential of the Stieltjes transform of the Gram matrix
$\sum_{n\neq m}\frac{q_nd_{m,n}}{N}\tilde{\mathbf{h}}_{m,n}\tilde{\mathbf{h}}_{m,n}^{\dag}$ and applying Theorem \ref{theorem_4}, the
denominator of (\ref{appen_eq_3}) converges almost surely to $-\phi_m'(\mathbf{q})\triangleq -\frac{\partial}{\partial
x}\phi_m(\mathbf{q},x)|_{x=w_m}$, which can be shown to be:
$\phi_m'(\mathbf{q})=\frac{-\phi_m(\mathbf{q})}{w_m+\frac{1}{N}\mathop{\sum}\limits_{n\neq
m}\frac{q_nd_{m,n}}{(1+q_nd_{m,n}\phi_m(\mathbf{q}))^2}}$.

The expression for $\frac{1}{N}|\tilde{\mathbf{h}}_{n,m}^{\dag}\mathbf{u}_n^*|^2$ can be further expanded as (29).

Following the same line of argument, the denominator of (\ref{appen_eq_4}) converges almost surely to $-\phi_n'(\mathbf{q})$. For the numerator
of (\ref{appen_eq_4}), since $\tilde{\mathbf{h}}_{n,m}$ and the Gram matrix $\sum_{\jmath\neq
n}\frac{q_{\jmath}d_{n,\jmath}}{N}\tilde{\mathbf{h}}_{n,\jmath}\tilde{\mathbf{h}}_{n,\jmath}^{\dag}$ are not independent, the numerator of
(\ref{appen_eq_4}) is transformed into the equivalent form (30) by matrix inversion lemma. 

By employing the rank-1 perturbation lemma \cite{silverstein95} and Theorem \ref{theorem_4}, the numerator of (\ref{appen_eq_5}) converges
almost surely to $-\phi_n'(\mathbf{q})$, and the denominator of (\ref{appen_eq_5}) converges almost surely to
$(1+q_m^*d_{n,m}\phi_n(\mathbf{q}))^2$. Combining the aforementioned results yields the fixed-point equation for
$\gamma_m^{\mathsf{PN}}(\mathbf{p})$ in (\ref{large_eq_9}). This completes the proof of Lemma \ref{lemma_3}.

\section*{Acknowledgment}
The authors would like to thank the anonymous reviewers and the Associate Editor for their suggestions which helped to improve the quality of
the paper.


\ifCLASSOPTIONcaptionsoff
  \newpage
\fi



%



%

\begin{biography}[{\includegraphics[width=1in,height=1.25in,clip,keepaspectratio]{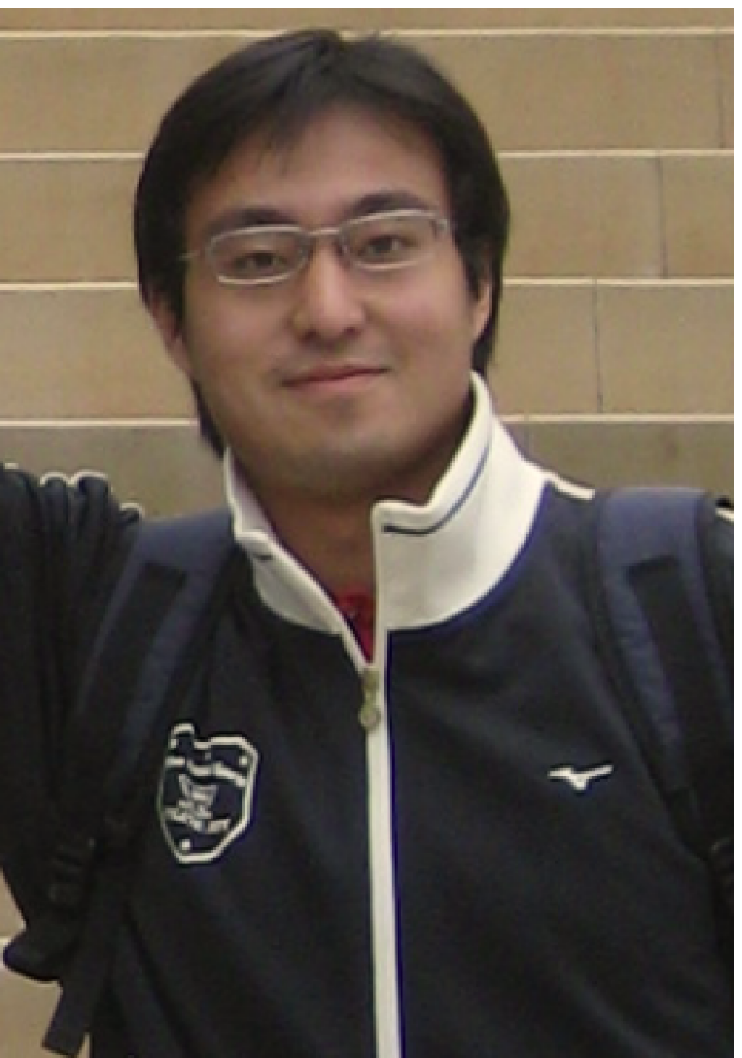}}]{Yichao Huang}
(S'10--M'12) received the B.Eng. degree in information engineering with highest honors from the Southeast University, Nanjing, China, in 2008,
and the M.S. and Ph.D. degrees in electrical engineering from the University of California, San Diego, La Jolla, in 2010 and 2012, respectively.
He then joined Qualcomm, Corporate R\&D, San Diego, CA, as a senior system engineer.

He interned with Qualcomm, Corporate R\&D, San Diego, CA, during summers 2011 and 2012. He was with California Institute for Telecommunications
and Information Technology (Calit2), San Diego, CA, during summer 2010. He was a visiting student at the Princeton University, Princeton, NJ,
during spring 2012. His research interests include communication theory, optimization theory, wireless networks, and signal processing for communication systems.

Dr. Huang received the President Fellowship both in 2005 and 2006, and the Best Thesis Award in 2008 from the Southeast University. He received the Microsoft Young Fellow Award in 2007 from Microsoft Research Asia and the ECE Department Fellowship from the University of California, San Diego in 2008 and was a finalist of Qualcomm Innovation Fellowship in 2010.
\end{biography}
\begin{biography}[{\includegraphics[width=1in,height=1.25in,clip,keepaspectratio]{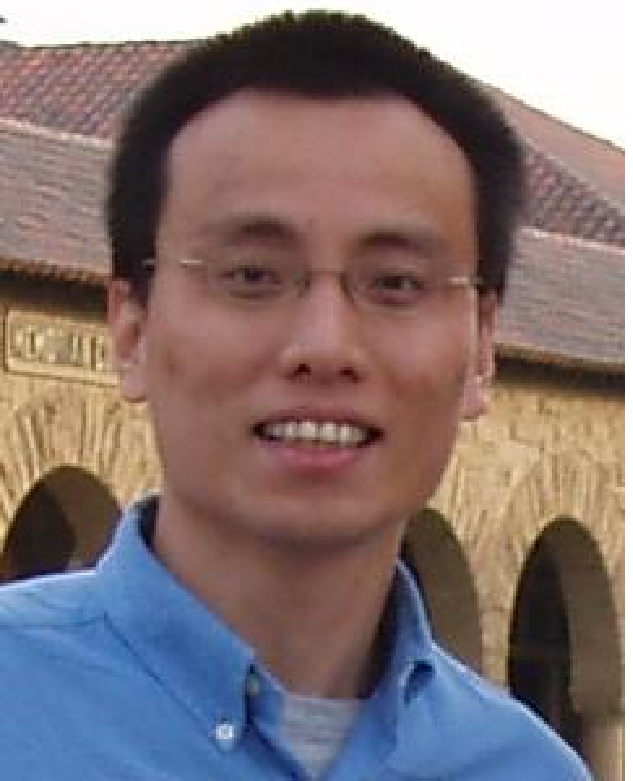}}]{Chee Wei Tan}
(M'08--SM'12) received the M.A. and Ph.D. degree in electrical engineering from Princeton University, Princeton, NJ, in 2006 and 2008, respectively.

He is an Assistant Professor at City University of Hong Kong. Previously, he was a Postdoctoral Scholar at the California Institute of Technology
(Caltech), Pasadena, CA. He was a Visiting Faculty at Qualcomm R\&D, San Diego, CA, in 2011. His research interests are in wireless and broadband
communications, complex networks, signal processing and nonlinear optimization.

Dr. Tan currently serves as an Editor for the IEEE TRANSACTIONS ON COMMUNICATIONS. He was the recipient of the 2008 Princeton University Wu Prize for Excellence and the 2011 IEEE Communications Society AP Outstanding Young Researcher Award. He was a selected participant at the U.S. National Academy of Engineering China-America Frontiers of Engineering Symposium in 2013.
\end{biography}
\begin{biography}[{\includegraphics[width=1in,height=1.25in,clip,keepaspectratio]{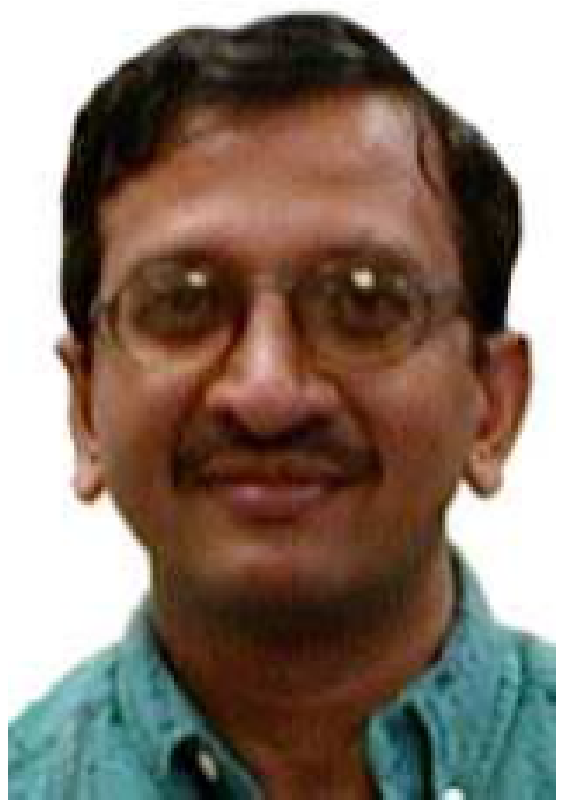}}]{Bhaskar D. Rao}
(S'80--M'83--SM'91--F'00) received the B.Tech. degree in electronics and electrical communication engineering from the Indian Institute of Technology, Kharagpur, India, in 1979 and the M.S. and Ph.D. degrees in electrical engineering from the University of Southern California, Los Angeles, in 1981 and 1983, respectively.

Since 1983, he has been with the University of California San Diego, La Jolla, where he is currently a Professor with the Electrical and Computer
Engineering Department and holder of the Ericsson endowed chair in wireless access networks. He was the Director of the Center for Wireless Communications from 2008-2011. His interests are in the areas of digital signal processing, estimation theory, and optimization theory, with applications to digital communications, speech signal processing, and human--computer interactions.

Dr. Rao was elected IEEE Fellow in 2000 for his contributions in high resolution spectral estimation. His research group has received several paper
awards. His paper received the best paper award at the 2000 Speech Coding Workshop and his students have received student paper awards at the 2005 and 2006 International Conference on Acoustics, Speech and Signal Processing as well as the best student paper award at the 2006 Neural Information Processing Systems (NIPS). A paper he co-authored with B. Song and R. Cruz received the 2008 Stephen O. Rice Prize Paper Award in the Field of Communications Systems. Another paper, he co-authored with David Wipf, received the 2012 signal processing society best paper award. He has been a member of the Statistical Signal and Array Processing technical committee, the Signal Processing Theory and Methods Technical Committee, the Communications Technical Committee of the IEEE Signal Processing Society. He has also served on the editorial board of the EURASIP Signal Processing Journal. He is a member of the signal processing society technical committee on machine learning for signal processing.
\end{biography}




\end{document}